\documentclass[]{aastex631}

\newcommand{\Lnmc}{L_{\rm NMC}}
\newcommand{\M}{M_{500}}

\newcommand{\Px}{\ensuremath{P_{\rm X}}}

\shorttitle{Looking for traces of non-minimally coupled DM}
\shortauthors{G. Gandolfi et al.}

\graphicspath{{./}{figures/}}

\usepackage{amsmath,latexsym,amssymb}
\usepackage{graphicx}
\usepackage{array}
\usepackage{cleveref}
\usepackage{caption}
\usepackage{subcaption}

\begin{document}

\nolinenumbers

\title{Looking for Traces of Non-minimally Coupled Dark Matter in the X-COP Galaxy Clusters Sample}

\author[0000-0003-3248-5666]{Giovanni Gandolfi}\affiliation{SISSA, Via Bonomea 265, 34136 Trieste, Italy}\affiliation{IFPU - Institute for fundamental physics of the Universe, Via Beirut 2, 34014 Trieste, Italy}\affiliation{INFN-Sezione di Trieste, via Valerio 2, 34127 Trieste, Italy}

\author[0000-0002-9153-1258]{Balakrishna S.Haridasu} \affiliation{SISSA, Via Bonomea 265, 34136 Trieste, Italy}\affiliation{IFPU - Institute for fundamental physics of the Universe, Via Beirut 2, 34014 Trieste, Italy}\affiliation{INFN-Sezione di Trieste, via Valerio 2, 34127 Trieste, Italy}

\author[0000-0002-7632-7443]{Stefano Liberati}\affiliation{SISSA, Via Bonomea 265, 34136 Trieste, Italy}\affiliation{IFPU - Institute for fundamental physics of the Universe, Via Beirut 2, 34014 Trieste, Italy}\affiliation{INFN-Sezione di Trieste, via Valerio 2, 34127 Trieste,  Italy}

\author[0000-0002-4882-1735]{Andrea Lapi}
\affiliation{SISSA, Via Bonomea 265, 34136 Trieste, Italy}\affiliation{IFPU - Institute for fundamental physics of the Universe, Via Beirut 2, 34014 Trieste, Italy}\affiliation{INFN-Sezione di Trieste, via Valerio 2, 34127 Trieste,  Italy}\affiliation{IRA-INAF, Via Gobetti 101, 40129 Bologna, Italy}

\begin{abstract}
We look for possible evidence of a non-minimal coupling (NMC) between dark matter (DM) and gravity using data from the X-COP compilation of galaxy clusters. We consider a theoretically motivated NMC that may dynamically arise from the collective behavior of the coarse-grained DM field (e.g., via Bose-Einstein condensation) with averaging/coherence length $\Lnmc$. In the Newtonian limit, the NMC modifies the Poisson equation by a term $\Lnmc^2 \nabla^2 \rho$ proportional to the Laplacian of the DM density itself. We show that this term when acting as a perturbation over the standard Navarro--Frenk--White (NFW) profile of cold DM particles, can yield DM halo density profiles capable of correctly fitting galaxy clusters' pressure profiles with an accuracy comparable and in some cases even better than the standard cold DM NFW profile. We also show that the observed relation between the non-minimal coupling length scale and the virial mass found in \cite{2022ApJ...929...48G} for Late Type Galaxies is consistent with the relation we find in the current work, suggesting that the previously determined power-law scaling law holds up to galaxy cluster mass scales.

\end{abstract}


\section{Introduction} \label{sec:intro}

\cite{1933AcHPh...6..110Z} originally hypothesized the existence of an unseen matter component to explain the large velocity scatter of the Coma cluster. In the subsequent decades, astrophysicists became aware of a discrepancy between luminous matter and the amount of mass required to explain the kinematic properties of spiral galaxies (\citealt{1978ApJ...225L.107R}; \citealt{1978PhDT.......195B}).
The astrophysical community traces back this missing mass to dark matter (DM); an unseen, cold (i.e. non-relativistic), and weakly interacting massive particle. This cold DM paradigm has been successful on cosmological scales, yet it struggles to fully reproduce the observed phenomenology on galactic scales, especially in DM-dominated dwarfs. This has motivated astrophysicists to consider several possible solutions, some of them radically departing from the standard cold DM paradigm. Some astrophysicists advocate a more realistic and complete inclusion of baryonic physics and feedback in models and simulations that could in principle alleviate some of the cold DM phenomenological issues at galactic scales (see e.g.
\citealt{DiCintio:2014xia};
\citealt{Pontzen:2014lma};
\citealt{El-Zant:2016byp}; 
\citealt{Navarro:2016bfs};
\citealt{2016MNRAS.455..476S};
\citealt{2017MNRAS.472.2153P};
\citealt{2017MNRAS.464.4160D};
\citealt{2017ApJ...835L..17K};
\citealt{2017PhRvL.118p1103L};
\citealt{Wheeler_2019};
\citealt{2020MNRAS.491.4523F}; 
\citealt{2020MNRAS.499.2912F}). Others point to alternative scenarios in which DM is composed of non-standard particle candidates (see the review by \citealt{2019A&ARv..27....2S} and references therein). Another proposal was to abandon entirely the DM paradigm in favour of modifying the laws of gravity (such as in the Modified Newtonian Dynamics or MOND model, originally proposed in \citealt{1983ApJ...270..365M}). 

In \cite{Gandolfi:2021jai} and \cite{2022ApJ...929...48G}, we have explored a new possibility to solve the small-scale incompleteness of the cold DM paradigm, having reviewed and tested a model in which cold DM is non-minimally coupled with gravity. Many other works by our team and collaborators already conjectured this possibility (e.g., \citealt{Bruneton:2008fk}; \citealt{Bertolami:2009ic}; \citealt{Bettoni:2011fs}; \citealt{Bettoni_2014}; \citealt{Bettoni_2015}; \citealt{Ivanov:2019iec}). As shown in \cite{Gandolfi:2021jai} and \cite{2022ApJ...929...48G}, the introduction of this coupling extends in a simple fashion the cold DM paradigm while maintaining its successful phenomenology on large cosmological scales and improving its behaviour in galactic systems. The term ``non-minimal'' implies that the gradient of the DM distribution directly couples to the Einstein tensor. Such non-minimal coupling (NMC) is not necessarily a fundamental feature of the DM particles, but rather may dynamically develop when the averaging/coherence length $ \Lnmc $ associated with the fluid description of the DM collective behaviour is comparable to the local curvature scale. In the Newtonian limit, this NMC appears as a modification of the Poisson equation by a term $ \Lnmc^2 \nabla^2 \rho $ proportional to the DM density $ \rho $ (see \citealt{Bettoni_2014}). This simple modification impacts on the internal dynamics of spiral galaxies, which are altered compared to a pure cold DM framework.
In \cite{Gandolfi:2021jai} and \cite{2022ApJ...929...48G} we have shown that this NMC between DM and gravity can alleviate the so-called core-cusp controversy, i.e. the observed discrepancy between the cored inner radii shape of the observed galactic dark haloes density profiles with the cuspier shape predicted by DM, gravity-only simulations who are best described by the so-called Navarro--Frenk--White (NFW) profile (\citealt{Navarro:1995iw}; \citealt{2001MNRAS.321..155L}; \citealt{Boylan-Kolchin:2003xvl}; \citealt{2006aglu.confE..30N}; \citealt{2010AdAst2010E...5D}; \citealt{Navarro:2016bfs}). \cite{2022ApJ...929...48G} also shown how such NMC manages to reproduce for a diverse sample of spiral galaxies the tight empirical relationships linking the baryonic and the dark component of galaxies. It is argued that the most general of such relations is the Radial Acceleration Relation (see \citealt{Lelli:2017vgz};  \citealt{Chae:2017bhk}; \citealt{Li:2018tdo};
\citealt{2018A&A...615A...3L}; \citealt{DiPaolo:2018mae}; \citealt{Green:2019cqm}; \citealt{Tian:2020qjd}; \citealt{rodrigues_marra_2020}), whose explanation is far from trivial in the cold DM framework (albeit some attempts have been made in this sense, see e.g. \citealt{DiCintio:2014xia}; \citealt{DiCintio:2015eeq}; \citealt{2016MNRAS.455..476S}; \citealt{2017ApJ...835L..17K}; \citealt{2017PhRvL.118p1103L}; \citealt{2017MNRAS.464.4160D}; \citealt{Navarro:2016bfs}; \citealt{Wheeler_2019}). 

The aim of the present work is to test the NMC DM model on the scales of galaxy clusters to assess its capability in fitting their pressure profiles and to determine if the scale relations predicted by this model are also satisfied in these regimes. For this purpose we will use the XMM-Newton Cluster Outskirts Project (X-COP) data products (see \citealt{2018A&A...614A...7G}; \citealt{2019A&A...621A..39E}; \citealt{2019A&A...621A..40E}; \citealt{2019A&A...621A..41G}). This sample consists of 12 clusters with well-observed X-ray emission and high signal to noise ratio in the Planck Sunyaev-Zel’dovich (SZ) survey (\citealt{2016A&A...594A..24P}). With the X-COP data we would have information about the ICM temperature and pressure in a wide radial range, from 0.2 Mpc to 2 Mpc. 

The paper is organized as follows: in Sec.~\ref{2|theory} we will briefly summarize the underlying theory behind the NMC DM model and we will present the data of the X-COP collaboration in more detail, in Sec.~\ref{3|results} we will proceed to illustrate and comment on our results and in Sec.~\ref{4|conclusion} we will summarize our work as well as outlining the future developments of our work.

Throughout this work, we adopt the standard flat $\Lambda$CDM cosmology (\citealt{Planck:2018vyg}) with rounded
parameter values: matter density $\Omega_{M}=0.3$, dark energy density $\Omega_{\Lambda}=0.7$, baryon density $\Omega_{b}=0.05$, and Hubble constant $H_{0}=100 h \mathrm{~km} \mathrm{~s}^{-1} \mathrm{Mpc}^{-1}$ with $h=0.7$. Unless otherwise specified, $G \approx 6.67 \times 10^{-8} \mathrm{~cm}^{3} \mathrm{~g}^{-1} \mathrm{~s}^{-2}$ indicates the standard gravitational (Newton) constant.

\section{NMC modelling and X-COP data}\label{2|theory}
\subsection{A theoretical background for the NMC}

Here we provide a short theoretical background for the NMC DM model, referring the reader to \cite{Gandolfi:2021jai} and \cite{2022ApJ...929...48G} for further information. A very basic NMC model can be built with the addition of a coupling term $S_{\rm int}$ between DM and gravity in the total Einstein--Hilbert action (in the Jordan frame) with shape:
\begin{equation}
\label{action}
    S_{\text {int }}\left[\tilde{g}_{\mu \nu}, \varphi\right] = \epsilon \Lnmc^{2} \int \mathrm{d}^{4} x \,\sqrt{-\tilde{g}} \, \widetilde{G}^{\mu \nu}\, \nabla_{\mu}\, \varphi \nabla_{\nu} \varphi~;
\end{equation}
here $\varphi$ is the (real) DM scalar field, $\epsilon = \pm 1$ is the polarity of the coupling, $\widetilde{G}^{\mu \nu}$ is the Einstein tensor, and $\Lnmc$ is the NMC characteristic length-scale. From a purely theoretical perspective, such form of the NMC is allowed by the Einstein equivalence principle (e.g., \citealt{Bekenstein:1992pj}; \citealt{2015AmJPh..83...39D}). In our approach however, the length $\Lnmc$ does not need to be a new fundamental constant of Nature, as it is indeed suggested by its virial mass-dependent scaling observed in \cite{2022ApJ...929...48G}. Instead, $\Lnmc$ could emerge dynamically from some collective behavior of the coarse-grained DM field (e.g., Bose-Einstein condensation). We hence remark that our NMC model does not consist in a modified gravity theory, but simply in a formalization of an emergent behavior of cold DM inside halos. Furthermore, the bookkeeping parameter $\epsilon$ will be set to $\epsilon=-1$ (repulsive coupling) based on the findings of \cite{Gandolfi:2021jai} and \cite{2022ApJ...929...48G}.

We also stress that the NMC DM model hereby discussed could in principle share features with other prospective DM models, such as self-interacting DM scenarios. Nonetheless, the NMC DM framework contemplates not only a self-interaction term for DM in the action but also a scale-dependent geometric interaction term between the DM field and the baryonic component, which is sourced by the non-minimal coupling of the DM to gravity.

Adopting the fluid approximation for the field $\varphi$ (as in \citealt{2012JCAP...07..027B}) and taking the Newtonian limit, the NMC translates into a simple modification of Poisson equation  (\citealt{Bettoni_2014})
\begin{equation}
\label{poissmod}
    \nabla^{2} \Phi=4 \pi G\left[\left(\rho+\rho_{\mathrm{bar}}\right)-\epsilon L^{2} \nabla^{2} \rho\right],
\end{equation}
where $\Phi$ is the Newtonian potential, and $\rho_{\mathrm{bar}}$ and $\rho$ are the baryonic and DM densities. In spherical symmetry, Eq.~(\ref{poissmod}) implies that the total gravitational acceleration writes 
\begin{equation}
\label{gnmc}
    g_{\text{tot}}(r) = -\frac{G\,M(< r)}{r^2}+4\pi\, G\, \epsilon L^2\, \frac{{\rm d}\rho}{{\rm d}r}\; ,
\end{equation}
where $M(<r)$ is the total mass enclosed in the radius $r$; the first term is the usual Newtonian acceleration and the second term is the additional contribution from the NMC.

In \cite{Gandolfi:2021jai} we have highlighted that Eq.~(\ref{poissmod}) gives rise to some interesting features for strongly DM-dominated systems in self-gravitating equilibria. First of all, the NMC can help to develop an inner core in the DM density profile. This enforces a shape for the density profile which closely follows the phenomenological Burkert profile (\citealt{1995ApJ...447L..25B}) out to several core scale radii. Moreover, DM-dominated halos with NMC are consistent with the core-column density relation (see e.g. \citealt{Salucci:2000ps}, \citealt{10.1111/j.1365-2966.2009.15004.x}, \citealt{burkert2015structure}, \citealt{2013ApJ...770...57B}, \citealt{Burkert:2020laq}), i.e. with the observed universality of the product between the core radius $r_0$ and the core density $\rho_0$. In \cite{2022ApJ...929...48G} we tested the NMC hypothesis using a diverse sample of spiral galaxies. The NMC DM model proved to yield fits to the stacked rotation curves of such objects with a precision always superior to pure NFW model fits and in several instances comparable or even better than the Burkert model ones. Furthermore, we observed an interesting power law scaling relation between the halo virial mass $M_{200}$ and the non-minimal coupling length scale $\Lnmc$ for the fitted galaxies. By assuming such mass-dependent scaling of $\Lnmc$, the NMC DM model was also able to reproduce the Radial Acceleration Relation up to the regime of dwarf spheroidal galaxies. Yet the NMC DM model awaits to be tested on scales larger than galactic ones, and this is precisely the scope of the present work.

\subsection{Modeling cluster thermal profiles}
The thermal pressure profiles\footnote{Here the gas density $n_{\rm gas} (r) \approx 1.826\, n_e (r)$ is the sum of the electron and proton number densities, $\mu$ is the mean molecular weight in a.m.u., and $m_p$ is the proton mass.} of galaxy clusters are defined as functions of the gravitational potential in play. In the framework of the NMC DM model this reads as

\begin{equation}
P^{\rm th}(R) = P^{\rm th}(0) - 1.8 \mu m_{\rm p}\int_0^R n_{\rm e}( r)\left[\frac{G {M}_{\rm DM}(r)}{r^2} - 4\pi\, G\, \epsilon \Lnmc^2\, \frac{{\rm d}\rho}{{\rm d}r} \right]{\rm d} r , 
\label{eq:pressure:profile:sz}
\end{equation}

where we model the electron density (ED) profile through the Vikhlinin profile \citep{Vikhlinin:2005mp}, 

\begin{equation}
\frac{n_e(r)}{n_0} = \frac{(r/r_c)^{-\alpha/2} [1 + (r/r_s)^{\gamma}]^{-\varepsilon/(2 \gamma)}}
{[1 + (r/r_c)^2]^{(3/2) \beta - \alpha/4}} .
\label{eq:vikgas}
\end{equation}

To specify the dark mass distribution in Eq.(\ref{eq:pressure:profile:sz}) we adopt the same perturbative approach of \cite{2022ApJ...929...48G}, considering the NMC as a small - perturbation over the standard cold DM NFW profile

\begin{equation}
\label{nfw}
    \rho_{\mathrm{NFW}}(r)=\frac{\delta_{\mathrm{c}} \rho_{\mathrm{c}} r_s^3}{r\left(r+r_s\right)^2}.
\end{equation}

Here, $r_s$ is a reference scale radius, $\delta_c$ is the dimensionless characteristic overdensity of the halo and $\rho_{\mathrm{c}}=3 H_0^2 / 8 \pi G$ is the local critical density. The NFW profile can also be written in terms of the halo virial mass $M_{500}$ (i.e., the mass value at which the interior mean density is 500 times the critical density of the Universe) and the halo concentration $c\equiv r_{500} / r_s$, with $r_{500} \approx 260\left(M_{500} / 10^{12} M_{\odot}\right)^{1 / 3}$ being the virial radius, and being $\delta_c \rho_c=M_{500} c^3 g(c) / 4 \pi r_{500}^3$  with $g(c) \equiv[\ln (1+c)-c /(1+c)]^{-1}$. The DM mass profile in Eq.(\ref{eq:pressure:profile:sz}) will then coincide with the NFW mass distribution, and the term $d\rho/dr$ will be the gradient of the NFW density profile. We remark that in this analysis the perturbative parameter is $\Lnmc/r_s$, a quantity that is always small for the range of masses probed in our study, as we will show with our results.


\subsection{The X-COP data}
We test the aforementioned formalism for the NMC using the XMM-Newton Cluster Outskirts Project (X-COP)\footnote{The datasets are publicly available at the following link: \href{https://dominiqueeckert.wixsite.com/xcop/about-x-cop}{https://dominiqueeckert.wixsite.com/xcop/about-x-cop}} catalogue \citep{Eckert:2016bfe} with joint X-ray temperature and Sunyaev--Zel'dovich (SZ) pressure observations. The methodology we adopt here is equivalent to the one earlier implemented in \citealt{Haridasu:2021hzq} (please refer to it for further details). To constrain the characteristic length scale ($\Lnmc$) alongside the parameters of the mass profile ($\Theta_{M}$) and the electron density (${\Theta}_e$), we write a joint likelihood $\cal{L}$ as


\begin{equation}
\label{eqn:likelihood}
\cal{L} = \cal{L}_{\rm P_X} + \cal{L}_{\rm P_{SZ}} + \cal{L}_{\rm ED},    
\end{equation}

where the pressure is computed through \cref{eq:pressure:profile:sz} and the electron density is modelled as \cref{eq:vikgas}. Here the first term accounts for the likelihood corresponding to the X-ray temperature $\Px$ data and the second term denotes the likelihood for the co-varying SZ pressure data and the last term in Eq.~(\ref{eqn:likelihood}) accounts for the modelled electron density data. 

Alongside these primary parameters of the model we also include an additional intrinsic scatter $\Sigma_{{\rm P, int}}$, following the approach in \cite{Ghirardini:2017apw, Ettori:2018tus}. We refer to \cite{Haridasu:2021hzq}, for an elaborate discussion on the mild differences between our approach here and the analysis performed in \cite{Ettori:2018tus}. 

We perform a Bayesian analysis through MCMC sampling using the publicly available \texttt{emcee}\footnote{\href{http://dfm.io/emcee/current/}{http://dfm.io/emcee/current/}} package \citep{Foreman-Mackey13, Hogg:2017akh}, which implements an affine-invariant ensemble sampler and \texttt{GetDist}\footnote{\href{https://getdist.readthedocs.io/}{https://getdist.readthedocs.io/}} package (\citealt{Lewis:2019xzd}), to perform analysis of the chains and plot the contours. We utilise flat uniform priors on all the parameters ${\Theta_{e}} = \{n_0, \alpha, \beta, \varepsilon, r_{\rm c}, r_{\rm s}\}$, $\Theta_{\rm M} = \{\M, c\}$ and the NMC characteristic length scale $\Lnmc$ in the MCMC analysis. Note here that we utilise the analytical form for the $M(<r)$ of the cluster which is expressed as a function of $\Theta_{\rm M}$. Finally, we also perform a model comparison through the Bayesian evidence $\cal{B}$ (\citealt{Trotta:2008qt, Trotta:2017wnx, Heavens:2017hkr}), using the \texttt{MCEvidence} package (\citealt{Heavens:2017afc})\footnote{\href{https://github.com/yabebalFantaye/MCEvidence}{https://github.com/yabebalFantaye/MCEvidence}.}. Comparing the Bayesian evidence one can assess the preference for a given model $\mathcal{M}_{1}(\Theta_{1})$ over the base model, i.e, the NFW model. Also, the Bayesian evidence is contrasted on the Jeffrey's scale \citep{Jeffreys:1939xee}, where $\Delta\log(\cal{B}) $ $ < 2.5 $ and $\Delta\log(\cal{B}) $ $> 5$, imply either a weak or a strong preference for the extended model, respectively.

\section{Testing the NMC with X-COP galaxy clusters data}
\label{3|results}

\subsection{General results and example clusters }
We report the results of our MCMC parameter estimation in Table~(\ref{fit}) and the respective statistical comparison in Table~(\ref{chisq}). The reduced chi-squared ($\chi^2_{\rm red}$) values in Table~(\ref{chisq}) indicate that for the majority of the clusters the NMC DM model generally provides a description of the data comparable and often even better than the NFW model. Nevertheless, we point out that the value of the NMC lengthscale $\Lnmc$ is partially guided by the availability of data at the innermost radii, and X-COP cluster pressure profiles are not well characterised in these regions. This lack of data at small radii relaxes the constraints on the higher-end of the possible values for $\Lnmc$, and it is ultimately responsible for the production of a hole-like feature (corresponding to low  or negative values of pressure) observed in our analysis for a certain fraction of the cluster pressure profiles at inner radii. We however anticipate that these features could be erased just by adding one or more data points at inner radii for the pressure profiles. Unfortunately, such data are yet to be available for the X-COP cluster sample. In light of this, the reader should interpret values of the NMC lengthscale $\Lnmc$ obtained in this work for clusters exhibiting a hole in their pressure profiles just as upper bounds on the real values of $\Lnmc$. We also note that our NMC DM model does not modify the estimation of pressure profiles in the outskirts of the cluster, essentially implying that the results presented here are not degenerate with any additional physics that can potentially affect the pressure profile estimation at outer radii, such as non-thermal pressure support, which for example could be important for cluster A2319 \citep{Eckert:2018mlz}. In the last column of Table~(\ref{chisq}) we show estimates of the Bayesian evidence $\Delta_{\mathcal{B}}$ exploited to further compare the two models, assuming standard NFW to be the base model. The NMC DM model is preferred for half of the clusters in the sample, and likewise it is mildly disfavored by the other half (up to the more striking case of RXC1825, for which $\Delta_{\mathcal{B}} = -3.53$). 

In Table~(\ref{fit}) we have reported the concentration $c$ and virial mass $M_{500}$ values from our MCMC analysis for the NFW and the NMC DM models. Estimates for these values from the two models are always compatible within the displayed uncertainties, with the exception of cluster RXC1825's concentration (slightly larger in the NMC framework than the NFW case) and $\M$ (conversely slightly smaller in the NMC case). Despite this overall compatibility, we note that the NMC model predicts concentration values systematically larger than the NFW ones. Table~(\ref{fit}) also features the MCMC estimations for the NMC lengthscale $\Lnmc$. Overall, these values of $\Lnmc$ exceed by two orders of magnitude on average the same values obtained for spiral galaxies in \cite{2022ApJ...929...48G}. This result is remarkably consistent with the increasing trend observed for spiral galaxies in \cite{2022ApJ...929...48G} between the mass of dark matter halos and the $\Lnmc$ associated with them, as we will show more in detail in Sect.~(\ref{l-mv}).  

\begin{figure}
    \centering
    \begin{subfigure}[b]{0.45\textwidth}
        \centering
        \includegraphics[width=1.25\textwidth]{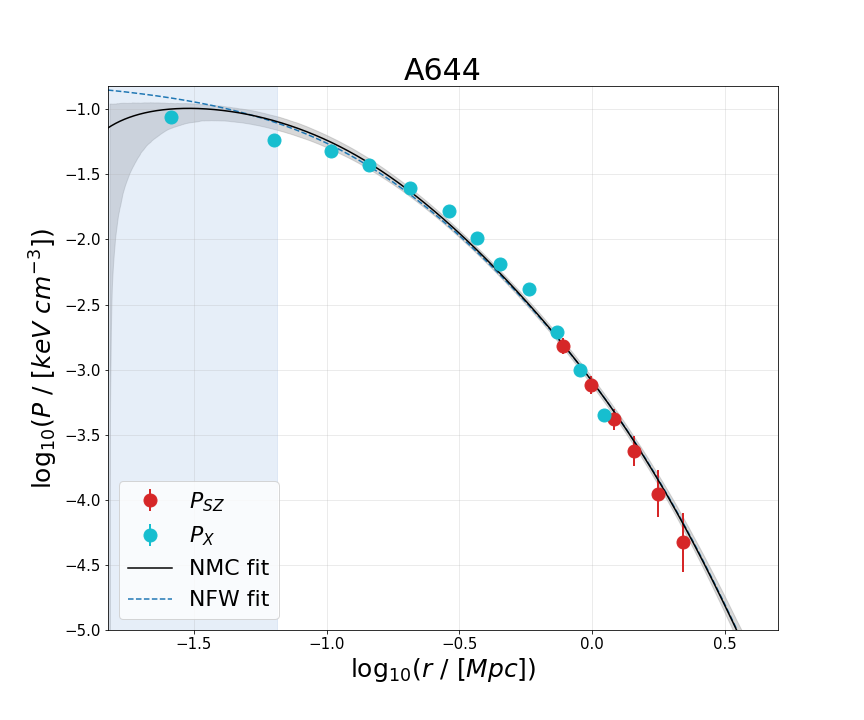}
        \label{fig:sfig1a}
    \end{subfigure}%
    \hfill
    \begin{subfigure}[b]{0.45\textwidth}
        \centering
        \includegraphics[width=.98\textwidth]{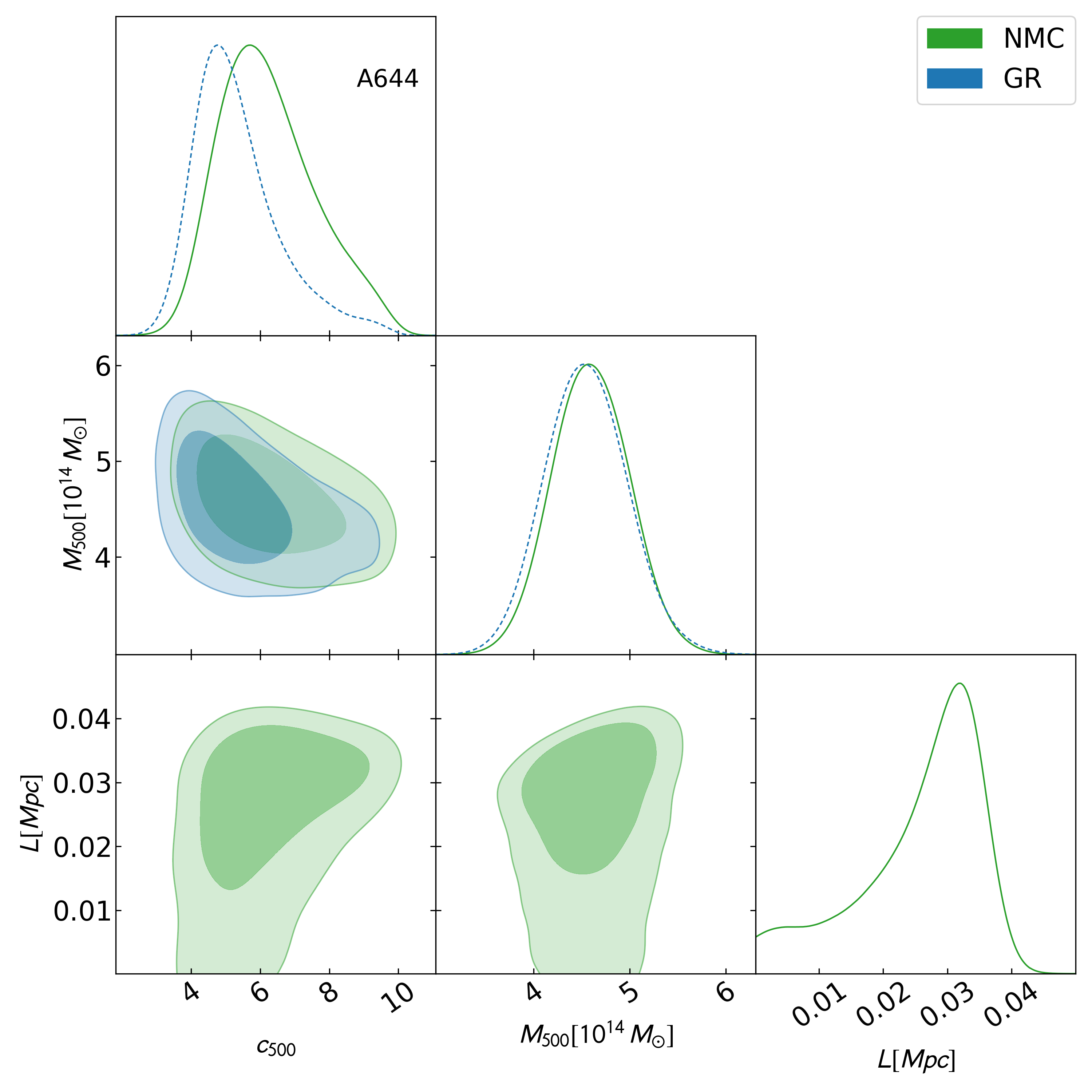}
        \label{fig:sfig1b}
    \end{subfigure}%
    \caption{Left: Pressure profile and related contour plots for the A644 cluster. Data are displayed as red dots (Sunyaev-Zel’dovich effect data) and cyan dots (data from the temperature profile by X-ray measurements). The black, solid lines represent the Bayesian MCMC best fit for the NMC DM model, with the grey contour representing the 68 \% confidence interval around the best fit line. The dashed blue line represents instead the NFW best fit. The blue shaded area in the profile represents the region of the dark halo within which the NMC is active, i.e. an area that extends from the centre of the halo up until $\Lnmc$. Right: The green contours represent the NMC DM model, while the blue contours represent the NFW fit.
    \label{profile1}}
\end{figure}

\begin{figure}
    \begin{subfigure}[b]{0.45\textwidth}
        \centering
        \includegraphics[width=1.25\textwidth]{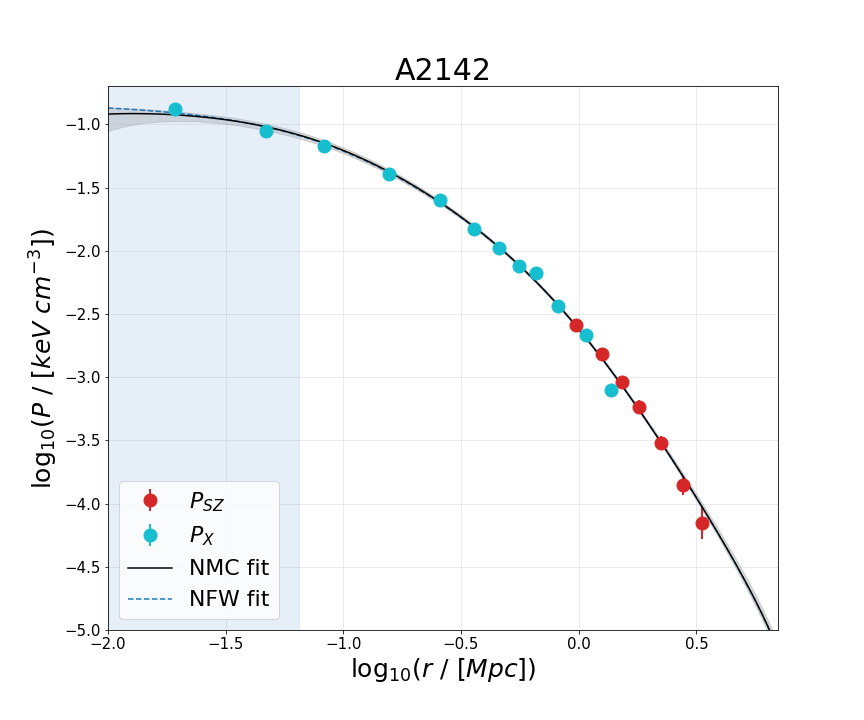}
        \label{fig:sfig2a}
    \end{subfigure}%
    \hfill
    \begin{subfigure}[b]{0.45\textwidth}
        \centering
        \includegraphics[width=0.98\textwidth]{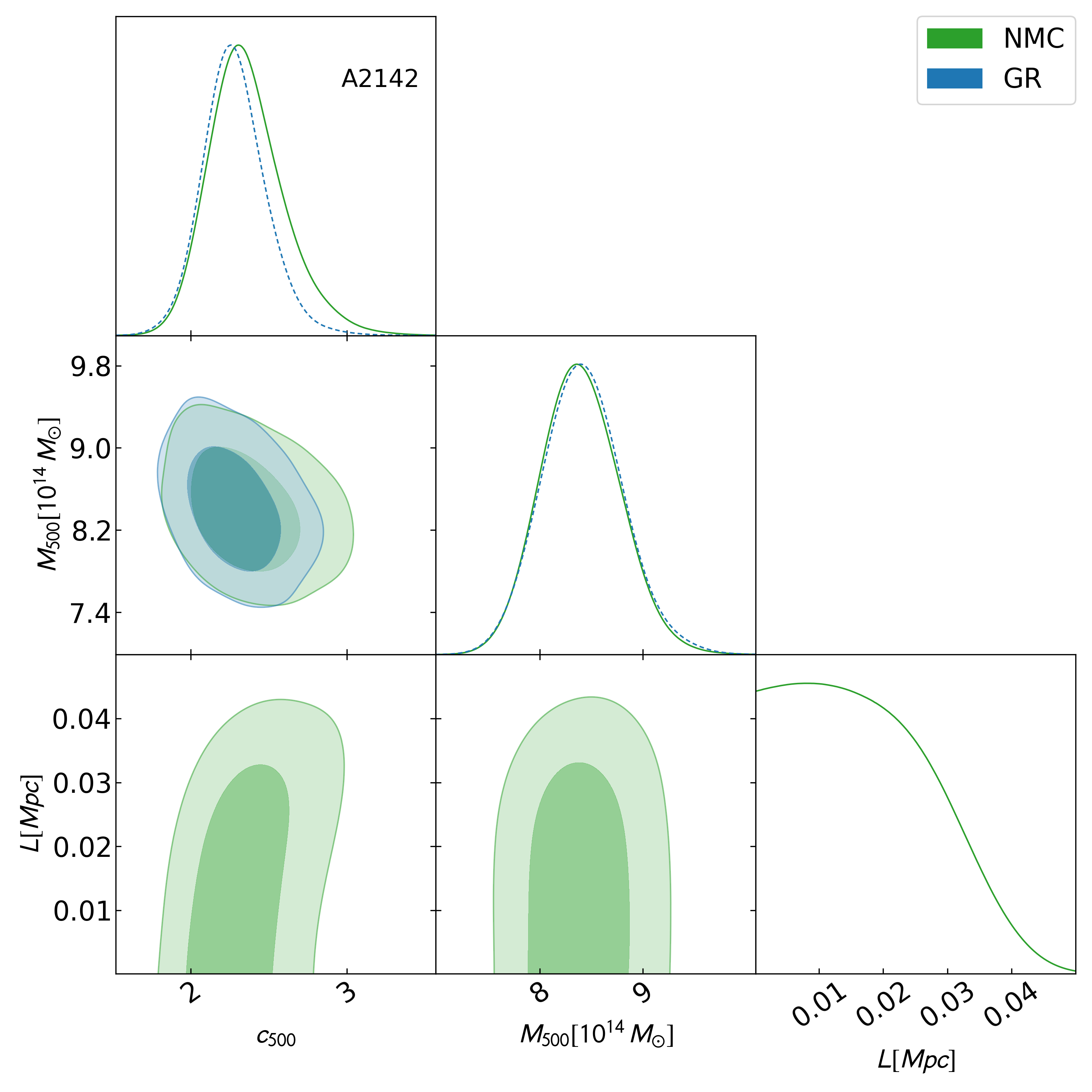}
        \label{fig:sfig2b}
    \end{subfigure}%
    \caption{Same as Fig.~(\ref{profile1}) but for the A2142 cluster. \label{profile2}}
\end{figure}

In Fig.~(\ref{profile1}) and Fig.~(\ref{profile2}) we show two exemplificative profiles (clusters A644 and A2142) obtained with our MCMC analysis, alongside the posterior contour plots for the $\{\M, c, \Lnmc\}$. As in the other clusters, both the NFW and the NMC DM models provide a good description of the general trend of the data. However, the NMC DM model is able to provide a better fit for the clusters whose data at the innermost radii are tracing a flattening in the shape of the pressure profiles. Such flattening seems to arise right within the area in which the NMC effect is active (i.e., within a distance of $\Lnmc$ from the center of the dark haloes, represented as a blue shaded area in both Fig.~(\ref{profile1}) and Fig.~(\ref{profile2})). As aforementioned, such NMC effect should be read with caution, given the limitation of the temperature data available in the innermost regions of the cluster.

\begin{figure}
    \centering
    \includegraphics[scale = .8]{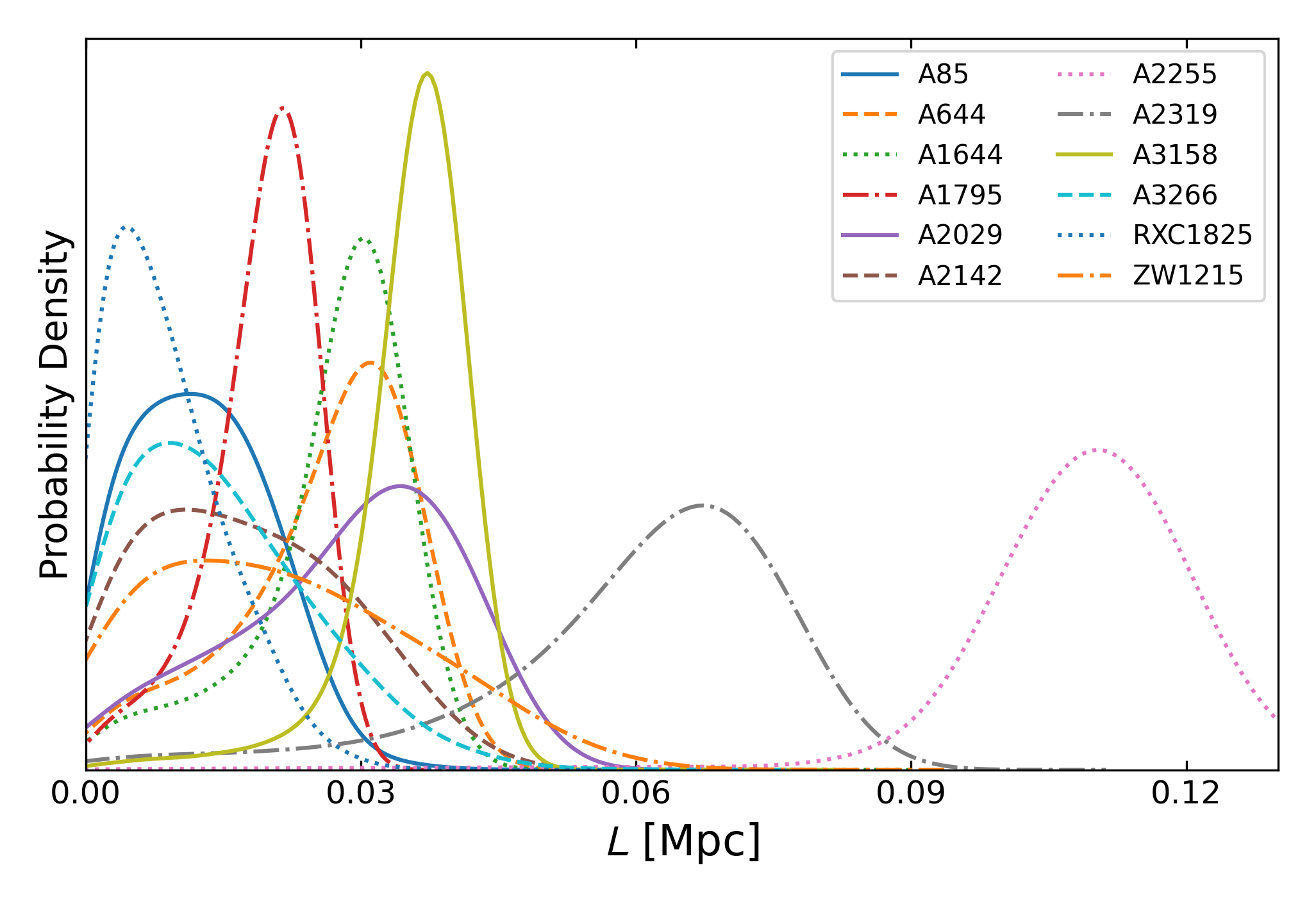}
    \caption{The one-dimensional posterior distribution for the lengthscale parameter $\Lnmc$ as retrieved in our Bayesian MCMC analysis..}
    \label{1dpost}
\end{figure}

Fig.~(\ref{1dpost}) shows the one-dimensional posterior distribution of the $\Lnmc$ parameter from our MCMC analysis for the X-COP cluster sample. Consistently with the galactic dark halos analyzed in \citealt{2022ApJ...929...48G}, $\Lnmc$ has different values in different halos, depending on their characteristics (in particular on their virial mass). Some halos (e.g. RXC1825 or A85) show a one-dimensional posterior converging towards $\Lnmc = 0$, suggesting that the dark matter density profile for these halos may have a cuspy shape, well reproduced by the NFW model. In other halos (e.g. A2319 and A2255) the NMC produces typical scale lengths capable of reaching fractions of Mpc. These values are likely to be slightly overestimated since, as previously discussed, some of these clusters exhibit an NMC DM pressure profile featuring a central hole. Despite this, the peak of such a one-dimensional posterior is clearly far from $\Lnmc = 0$, indicating that the shape of the density profile of these dark halos could be less cuspy and different from that of the NFW profile. As can be seen in the right panel of Fig.~(\ref{profile1}), the non-zero values for $\Lnmc$ are essentially accompanied by a mild positive correlation with $\M$ and subsequently a non-Gaussian degeneracy with the concentration $c$. Also, for all the clusters that have a non-zero posterior for the $\Lnmc$, we do not observe any such correlation with the $\M$ parameter, as in the case of A2142, shown in the right panel of Fig.~(\ref{profile2}). In this context, clusters A2255 and A2319 show a slightly larger value of the lengthscale $\Lnmc$ in the posteriors. We also note that for the clusters A2255 and RXC1825, we find a strong bi-modal behavior, from which we select the maximum posterior region.
As can be seen also from the corresponding Bayesian evidence in favor of the NMC DM model, the clusters A3158, A2319, and A2255 show a moderate preference ($\Delta_{log(\mathcal{B})}\gtrsim 2$), owing to the slightly larger values of $\Lnmc$. As can be seen in \Cref{fig:profiles_all}, this evidence in favor of the NMC DM in these three clusters is essentially driven by the improvement of the fit accounting for the innermost data point in the X-ray pressure observations. And on the contrary, the cluster RXC1825 shows a preference for the standard NFW scenario at a similar level of Bayesian evidence.

\subsection{$\Lnmc$ vs. $\M$}
\label{l-mv}

\begin{figure}
    \centering
    \includegraphics[scale = 0.5]{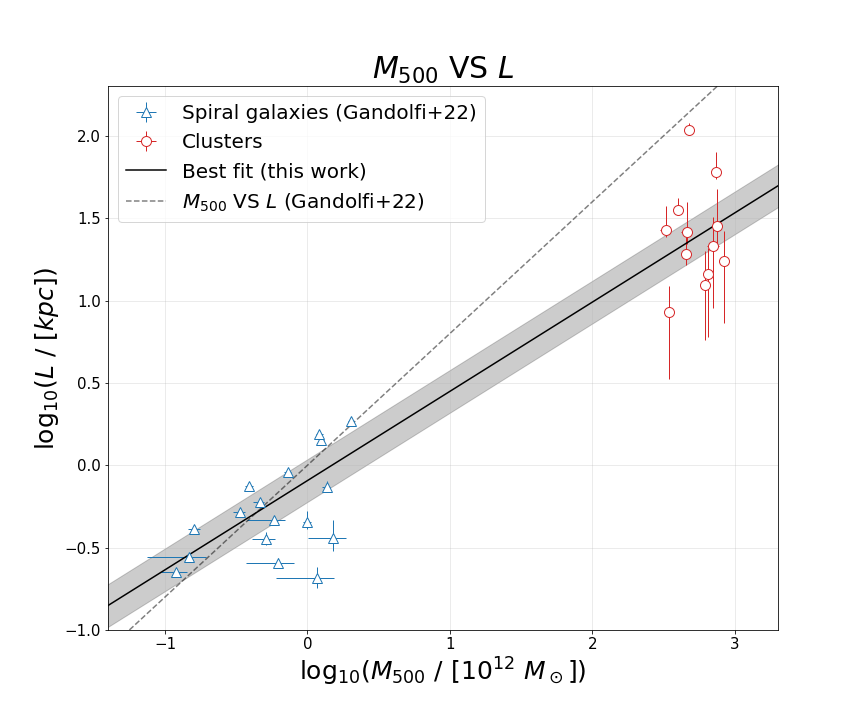}
    \caption{Virial mass ($M_{500}$) vs. $\Lnmc$ relation. Blue triangles are the same spiral galaxies data utilized in \citealt{2022ApJ...929...48G}, whereas the red circles represent the X-COP cluster measurements found in our Bayesian MCMC analysis. The best-fit power law is represented as a black solid line, whereas the shaded grey area represents a one-sigma confidence interval. The grey dashed line represents the $M_{500}$ VS $\Lnmc$ relation utilized in \citealt{2022ApJ...929...48G} to obtain the results therein. Note that the virial masses of spirals and their errors are rescaled to $M_{500}$ (i.e., a mass at which the interior mean density is 500 times the critical density of the Universe) since they were originally computed as $M_{200}$ (i.e., a mass at which the interior mean density is 200 times the critical density of the Universe).}
    \label{figmv-lnmc}
\end{figure}

In this section, we investigate the relation between the NMC lengthscale $\Lnmc$ and the dark halo virial mass $M_{500}$ observed as a result of our analysis. We remark that this relationship is an important feature of the NMC DM model which, as previously stated, is not to be considered as a modified theory of gravity, and therefore $\Lnmc$ should not be thought of as a new proposed fundamental constant of nature. The observed relationship between $\Lnmc$ and $M_{500}$ shows that $\Lnmc$ indeed does not have a universal value, and it depends on at least one property of the dark haloes under consideration.
The $\Lnmc$ - $M_{500}$ relationship was first observed in \cite{2022ApJ...929...48G} to hold for the galactic dark halos, analyzed therein. A remarkable result of this earlier analysis is that one can describe such a relationship with a simple power law. In this work, we investigate the validity of this relation up to the virial mass ranges typical of galaxy clusters. The results of our analysis are shown in Fig.~(\ref{figmv-lnmc}). Here, the virial masses of the spiral galaxies and their errors are rescaled from $M_{200}$ to $M_{500}$ to homogenize the results. Remarkably, the X-COP clusters data point derived by our MCMC analysis are seemingly in agreement with the power law trend of the $\Lnmc$ - $M_{500}$ relationship observed in \cite{2022ApJ...929...48G}. We performed an MCMC fit using the model $\log_{10}\Lnmc = a \log_{10} (b M_{500})$ to fit both galactic and clusters data simultaneously, obtaining as parameter values $a=0.542 \pm 0.005$ and $b=0.807 \pm 0.005$. The slope $a$ found in this analysis is compatible with the slope found by fitting a similar power law to galaxies only, as done in \citealt{2022ApJ...929...48G} ($0.7 \pm 0.2$). The best fit line in this work is shown in Fig.~(\ref{figmv-lnmc}) as a solid black line together with a grey shaded area representing a one-sigma confidence limit of the fit. In the same figure, we also show as a grey dotted line the relation $\Lnmc = M_{200}^{0.8}$, utilized in \citealt{2022ApJ...929...48G} as a reference relation to study the capacity of the NMC DM model in reproducing the Radial Acceleration Relation. In the galactic virial mass regime, the two power laws are consistent within a one-sigma confidence limit, and their slopes are compatible within the errors. The updated scaling law retrieved in this work translates into an average variation of the RAR with respect to the one computed in \citealt{2022ApJ...929...48G} by a mere 0.33$\%$, with the average of such variation being taken for every radial acceleration bin in which the RAR of \citealt{2022ApJ...929...48G} is computed (spanning from a minimum variation of 0.004$\%$ to 1.4$\%$ among all the bins). We stress that such variation is well within the errors associated to the RAR computed in \citealt{2022ApJ...929...48G} for every single bin of radial acceleration. In fact, for the RAR of \citealt{2022ApJ...929...48G} the minimum and maximum percentage relative uncertainties are 0.67$\%$ and 3.27$\%$ respectively, and the average one is 1.85$\%$. We thus conclude that the updated $\Lnmc$ - $M_{500}$ relation retrieved in this work, albeit different from the one considered in \citealt{2022ApJ...929...48G}, is still able to reproduce the RAR in the galactic dark haloes mass regime. That being said, from Fig.~(\ref{figmv-lnmc}) it is possible to appreciate the significant difference between the two power laws when approaching the cluster dark halo mass regime. This essentially constitutes an improvement over the previous analysis which utilized only the galaxies to assess the same relation. As previously mentioned, for some of the clusters the $\Lnmc$ values could be slightly overestimated, and hence it is possible that the real best-fit power law could be even less steep than what is found in our analysis. Moreover, we expect that including a galaxy cluster dataset that probes the innermost regions of the halo could help reduce the scatter in the $\Lnmc$ - $M_{500}$ relation.
  
\subsection{Scatter in the $\M$ vs. $c$ }

\begin{figure}
    \centering
    \includegraphics[scale = 0.5]{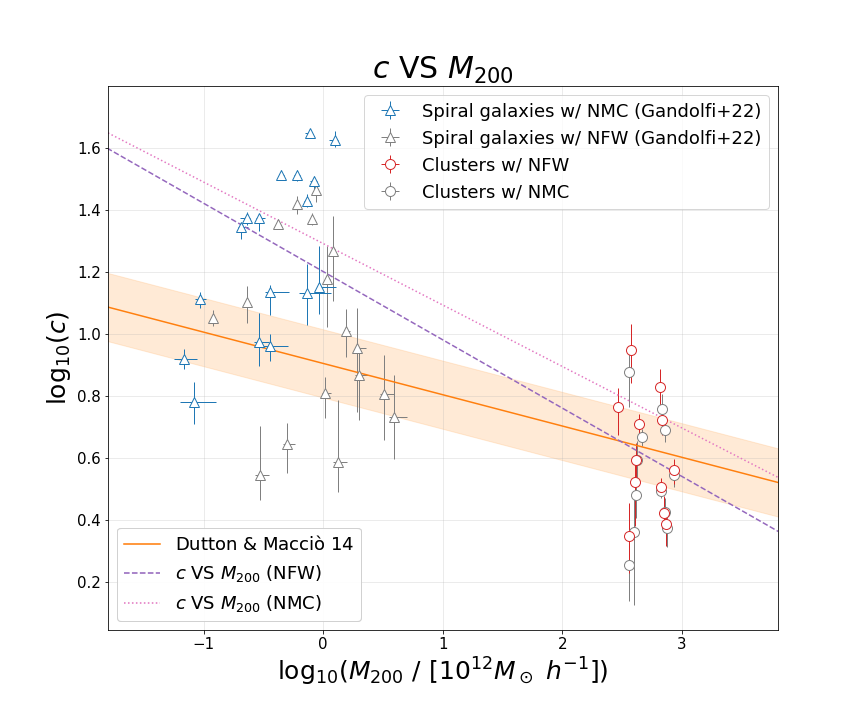}
    \caption{Concentration VS virial mass relation. Grey triangles and grey circles are respectively spiral galaxies from \citealt{2022ApJ...929...48G} and X-COP clusters' data obtained with the NFW model. Blue triangles and red circles represent data retrieved assuming the NMC DM model. The orange solid line represents the relation by \citealt{Dutton:2014xda} featuring a lognormal scatter of 0.11 dex represented by the orange area around the line. The purple dashed line and the pink dashed lines represent respectively the $c_{200}$ VS $M_{200}$ relations respectively found for the NFW model and the NMC DM model. Note that the cluster virial masses ($M_{500}$) and their errors have been downscaled to $M_{200}$ to make them comparable to the \citealt{Dutton:2014xda} relation.}
    \label{c-mv}
\end{figure}

In Fig.~(\ref{c-mv}) we test the correlation between concentration $c$ and $M_ {500}$ values inferred from our MCMC analysis against the relationship between $c_{200}$ and $M_{200}$ of dark halos found in \citealt{Dutton:2014xda}, namely:

\begin{equation}
    \log_{10} c_{200}=0.905-0.101 \log_{10} \left(M_{200} / 10^{12} h^{-1} M_{\odot}\right).
\end{equation}

To make this comparison, we rescale the value of the virial mass $M_{500}$ of the clusters to $M_{200}$, recalculating the corresponding concentrations accordingly. We then perform an MCMC fit to find the best-fit power law that best describes the data obtained by exploiting both the NFW model and the NMC DM model. In both these cases, there is some visible difference between the two best-fit power laws and the relationship found in \citealt{Dutton:2014xda}. This is true at least up to the cluster mass regime, where both the best-fit power laws of the NFW and NMC DM model intersect the report of \citealt{Dutton:2014xda}. Comparing the best-fit power laws with each other, we do not identify important differences between the two models, since the corresponding data have a rather similar scatter around the \citealt{Dutton:2014xda} relation. This is something we expected following the previous examination of the tabulated results of our MCMC analysis. \Cref{c-mv} can provide interesting qualitative hints on the expected concentrations of sub-haloes in galaxy clusters within this framework. As shown in \citealt{2020Sci...369.1347M}, the $\Lambda$CDM is at variance with the observed density and compactness of the dark matter sub-haloes in galaxy clusters. From our analysis, the NMC DM model predicts galaxy-sized dark matter sub-structures in clusters featuring overall higher concentrations associated with lower halo mass values with respect to the standard CDM paradigm. However, we caveat that only future analysis relying on high-quality data and exploiting a larger sample of galaxy clusters can confirm this prediction. In this context, the observed tensions at galaxy clusters scales present a promising way to further test the NMC dark matter scenario and its phenomenology.

\section{Summary}\label{4|conclusion}

In this section, we summarize the main results of this work. We tested the NMC DM model against the pressure profiles of galaxy clusters belonging to the X-COP sample, finding that:

\begin{itemize}
\item Our model in which the NMC act as a perturbation over a cold DM behavior provides a good description of the cluster pressure profiles, with a fit accuracy comparable to or in some cases even better than the NFW model both in terms of reduced $\chi^2$ and Bayesian evidence;
\item The $ M_ {500} - \Lnmc $ relation is well described by a simple power law even beyond the mass regime of spiral galaxies investigated in \citealt{2022ApJ...929...48G}. However, wanting to extend this relationship to include galaxy clusters, it is necessary to correct the slope of the above relationship with respect to the value reported in \citealt{2022ApJ...929...48G} based only on LTGs.
\end{itemize}

One key issue in our analysis is the lack of data at smaller radii in the pressure profiles of the X-COP clusters, as this may have partially resulted in overestimating the $ \Lnmc $ values inferred in our analysis. Nevertheless, previous works based on X-COP cluster data (see e.g. \citealt{Haridasu:2021hzq}) highlighted how cored profiles would seem to better describe the DM density distribution for a few clusters belonging to this sample. Then, even if the X-COP cluster profiles were better characterized at inner radii, the NMC DM model would probably be still preferred for all those clusters exhibiting cored profiles with respect to the cuspier NFW model. Indeed, a possible future step to corroborate our analysis would be to use data from well-characterized galaxy clusters at small radii (such as data from the CLASH collaboration, see e.g. \citealt{2014ApJ...795..163U}), probing the regions where the effect of the NMC is crucial. Another interesting extension of our work would concern the investigation of the mechanism originating the NMC between DM and gravity, and particularly how this mechanism gives rise to the observed power law relationship between $ \Lnmc $ and the virial mass $ M_ {500} $. For this purpose, we will consider implementing the NMC DM model in full N-body simulation to study the time-dependent conditions and the formation mechanisms of cosmic structures in this framework. In this context, colliding galaxy clusters would configure as promising study systems to place constraints on the NMC DM model, in a similar fashion to what is done with self-interacting DM scenarios (see e.g. \citealt{Robertson:2016xjh}). Indeed, the effects of the NMC in colliding systems could be particularly significant in the regions where the DM density changes appreciably as a consequence of the DM haloes merger. Indeed, we expect the repulsive nature of the NMC to manifest at the interface of the collision, with the overall effect of slowing down the merger process. Modelizing this scenario is however challenging and calls for a dedicated future work. We also stress that another interesting avenue to characterize the phenomenology of this model further is to test it against known tensions on galaxy cluster scales and beyond.
\\
\\
We warmly thank Dominique Eckert for sharing additional data with us, and we thank the anonymous referee for the helpful and constructive comments. AL is supported by the EU H2020-MSCA-ITN-2019 Project 860744 BiD4BESt: Big Data applications for black hole Evolution Studies, and by the PRIN MIUR 2017 prot. 20173ML3WW: Opening the ALMA window on the cosmic evolution of gas, stars and supermassive black holes. BSH is supported by the INFN INDARK grant.

\clearpage

\bibliography{bibliography}{}

\begin{thebibliography}{}
\expandafter\ifx\csname natexlab\endcsname\relax\def\natexlab#1{#1}\fi
\providecommand{\url}[1]{\href{#1}{#1}}
\providecommand{\dodoi}[1]{doi:~\href{http://doi.org/#1}{\nolinkurl{#1}}}
\providecommand{\doeprint}[1]{\href{http://ascl.net/#1}{\nolinkurl{http://ascl.net/#1}}}
\providecommand{\doarXiv}[1]{\href{https://arxiv.org/abs/#1}{\nolinkurl{https://arxiv.org/abs/#1}}}

\bibitem[{Aghanim {et~al.}(2020)}]{Planck:2018vyg}
Aghanim, N., {et~al.} 2020, Astron. Astrophys., 641, A6,
  \dodoi{10.1051/0004-6361/201833910}

\bibitem[{{Behroozi} {et~al.}(2013){Behroozi}, {Wechsler}, \&
  {Conroy}}]{2013ApJ...770...57B}
{Behroozi}, P.~S., {Wechsler}, R.~H., \& {Conroy}, C. 2013, \apj, 770, 57,
  \dodoi{10.1088/0004-637X/770/1/57}

\bibitem[{Bekenstein(1993)}]{Bekenstein:1992pj}
Bekenstein, J.~D. 1993, Phys. Rev. D, 48, 3641,
  \dodoi{10.1103/PhysRevD.48.3641}

\bibitem[{Bertolami \& Paramos(2010)}]{Bertolami:2009ic}
Bertolami, O., \& Paramos, J. 2010, JCAP, 03, 009,
  \dodoi{10.1088/1475-7516/2010/03/009}

\bibitem[{Bettoni {et~al.}(2014)Bettoni, Colombo, \& Liberati}]{Bettoni_2014}
Bettoni, D., Colombo, M., \& Liberati, S. 2014, Journal of Cosmology and
  Astroparticle Physics, 2014, 004–004, \dodoi{10.1088/1475-7516/2014/02/004}

\bibitem[{Bettoni \& Liberati(2015)}]{Bettoni_2015}
Bettoni, D., \& Liberati, S. 2015, Journal of Cosmology and Astroparticle
  Physics, 2015, 023, \dodoi{10.1088/1475-7516/2015/08/023}

\bibitem[{Bettoni {et~al.}(2011)Bettoni, Liberati, \& Sindoni}]{Bettoni:2011fs}
Bettoni, D., Liberati, S., \& Sindoni, L. 2011, JCAP, 11, 007,
  \dodoi{10.1088/1475-7516/2011/11/007}

\bibitem[{{Bettoni} {et~al.}(2012){Bettoni}, {Pettorino}, {Liberati}, \&
  {Baccigalupi}}]{2012JCAP...07..027B}
{Bettoni}, D., {Pettorino}, V., {Liberati}, S., \& {Baccigalupi}, C. 2012,
  \jcap, 2012, 027, \dodoi{10.1088/1475-7516/2012/07/027}

\bibitem[{{Bosma}(1978)}]{1978PhDT.......195B}
{Bosma}, A. 1978, PhD thesis, -

\bibitem[{Boylan-Kolchin \& Ma(2004)}]{Boylan-Kolchin:2003xvl}
Boylan-Kolchin, M., \& Ma, C.-P. 2004, Mon. Not. Roy. Astron. Soc., 349, 1117,
  \dodoi{10.1111/j.1365-2966.2004.07585.x}

\bibitem[{Bruneton {et~al.}(2009)Bruneton, Liberati, Sindoni, \&
  Famaey}]{Bruneton:2008fk}
Bruneton, J.-P., Liberati, S., Sindoni, L., \& Famaey, B. 2009, JCAP, 03, 021,
  \dodoi{10.1088/1475-7516/2009/03/021}

\bibitem[{{Burkert}(1995)}]{1995ApJ...447L..25B}
{Burkert}, A. 1995, \apjl, 447, L25, \dodoi{10.1086/309560}

\bibitem[{Burkert(2015)}]{burkert2015structure}
Burkert, A. 2015, The Structure and Dark Halo Core Properties of Dwarf
  Spheroidal Galaxies.
\newblock \doarXiv{1501.06604}

\bibitem[{Burkert(2020)}]{Burkert:2020laq}
---. 2020, Astrophys. J., 904, 161, \dodoi{10.3847/1538-4357/abb242}

\bibitem[{Chae {et~al.}(2019)Chae, Bernardi, Sheth, \& Gong}]{Chae:2017bhk}
Chae, K.-H., Bernardi, M., Sheth, R.~K., \& Gong, I.-T. 2019, Astrophys. J.,
  877, 18, \dodoi{10.3847/1538-4357/ab18f8}

\bibitem[{{de Blok}(2010)}]{2010AdAst2010E...5D}
{de Blok}, W.~J.~G. 2010, Advances in Astronomy, 2010, 789293,
  \dodoi{10.1155/2010/789293}

\bibitem[{{Desmond}(2017)}]{2017MNRAS.464.4160D}
{Desmond}, H. 2017, \mnras, 464, 4160, \dodoi{10.1093/mnras/stw2571}

\bibitem[{{Di Casola} {et~al.}(2015){Di Casola}, {Liberati}, \&
  {Sonego}}]{2015AmJPh..83...39D}
{Di Casola}, E., {Liberati}, S., \& {Sonego}, S. 2015, American Journal of
  Physics, 83, 39, \dodoi{10.1119/1.4895342}

\bibitem[{Di~Cintio {et~al.}(2014)Di~Cintio, Brook, Dutton, Macci\`o, Stinson,
  \& Knebe}]{DiCintio:2014xia}
Di~Cintio, A., Brook, C.~B., Dutton, A.~A., {et~al.} 2014, Mon. Not. Roy.
  Astron. Soc., 441, 2986, \dodoi{10.1093/mnras/stu729}

\bibitem[{Di~Cintio \& Lelli(2016)}]{DiCintio:2015eeq}
Di~Cintio, A., \& Lelli, F. 2016, Mon. Not. Roy. Astron. Soc., 456, L127,
  \dodoi{10.1093/mnrasl/slv185}

\bibitem[{Di~Paolo {et~al.}(2019)Di~Paolo, Salucci, \&
  Fontaine}]{DiPaolo:2018mae}
Di~Paolo, C., Salucci, P., \& Fontaine, J.~P. 2019, Astrophys. J., 873, 106,
  \dodoi{10.3847/1538-4357/aaffd6}

\bibitem[{Donato {et~al.}(2009)Donato, Gentile, Salucci, Frigerio~Martins,
  Wilkinson, Gilmore, Grebel, Koch, \& Wyse}]{10.1111/j.1365-2966.2009.15004.x}
Donato, F., Gentile, G., Salucci, P., {et~al.} 2009, Monthly Notices of the
  Royal Astronomical Society, 397, 1169,
  \dodoi{10.1111/j.1365-2966.2009.15004.x}

\bibitem[{Dutton \& Macci\`o(2014)}]{Dutton:2014xda}
Dutton, A.~A., \& Macci\`o, A.~V. 2014, Mon. Not. Roy. Astron. Soc., 441, 3359,
  \dodoi{10.1093/mnras/stu742}

\bibitem[{Eckert {et~al.}(2017)Eckert, Ettori, Pointecouteau, Molendi, Paltani,
  \& Tchernin}]{Eckert:2016bfe}
Eckert, D., Ettori, S., Pointecouteau, E., {et~al.} 2017, Astron. Nachr., 338,
  293, \dodoi{10.1002/asna.201713345}

\bibitem[{{Eckert} {et~al.}(2019){Eckert}, {Ghirardini}, {Ettori}, {Rasia},
  {Biffi}, {Pointecouteau}, {Rossetti}, {Molendi}, {Vazza}, {Gastaldello},
  {Gaspari}, {De Grandi}, {Ghizzardi}, {Bourdin}, {Tchernin}, \&
  {Roncarelli}}]{2019A&A...621A..40E}
{Eckert}, D., {Ghirardini}, V., {Ettori}, S., {et~al.} 2019, \aap, 621, A40,
  \dodoi{10.1051/0004-6361/201833324}

\bibitem[{Eckert {et~al.}(2019)}]{Eckert:2018mlz}
Eckert, D., {et~al.} 2019, Astron. Astrophys., 621, A40,
  \dodoi{10.1051/0004-6361/201833324}

\bibitem[{El-Zant {et~al.}(2016)El-Zant, Freundlich, \&
  Combes}]{El-Zant:2016byp}
El-Zant, A., Freundlich, J., \& Combes, F. 2016, Mon. Not. Roy. Astron. Soc.,
  461, 1745, \dodoi{10.1093/mnras/stw1398}

\bibitem[{{Ettori} {et~al.}(2019){Ettori}, {Ghirardini}, {Eckert},
  {Pointecouteau}, {Gastaldello}, {Sereno}, {Gaspari}, {Ghizzardi},
  {Roncarelli}, \& {Rossetti}}]{2019A&A...621A..39E}
{Ettori}, S., {Ghirardini}, V., {Eckert}, D., {et~al.} 2019, \aap, 621, A39,
  \dodoi{10.1051/0004-6361/201833323}

\bibitem[{Ettori {et~al.}(2019)Ettori, Ghirardini, Eckert, Pointecouteau,
  Gastaldello, Sereno, Gaspari, Ghizzardi, Roncarelli, \&
  Rossetti}]{Ettori:2018tus}
Ettori, S., Ghirardini, V., Eckert, D., {et~al.} 2019, Astron. Astrophys., 621,
  A39, \dodoi{10.1051/0004-6361/201833323}

\bibitem[{{Foreman-Mackey} {et~al.}(2013){Foreman-Mackey}, {Hogg}, {Lang}, \&
  {Goodman}}]{Foreman-Mackey13}
{Foreman-Mackey}, D., {Hogg}, D.~W., {Lang}, D., \& {Goodman}, J. 2013,
  Publications of the Astronomical Society of the Pacific, 125, 306,
  \dodoi{10.1086/670067}

\bibitem[{{Freundlich} {et~al.}(2020{\natexlab{a}}){Freundlich}, {Dekel},
  {Jiang}, {Ishai}, {Cornuault}, {Lapiner}, {Dutton}, \&
  {Macci{\`o}}}]{2020MNRAS.491.4523F}
{Freundlich}, J., {Dekel}, A., {Jiang}, F., {et~al.} 2020{\natexlab{a}},
  \mnras, 491, 4523, \dodoi{10.1093/mnras/stz3306}

\bibitem[{{Freundlich} {et~al.}(2020{\natexlab{b}}){Freundlich}, {Jiang},
  {Dekel}, {Cornuault}, {Ginzburg}, {Koskas}, {Lapiner}, {Dutton}, \&
  {Macci{\`o}}}]{2020MNRAS.499.2912F}
{Freundlich}, J., {Jiang}, F., {Dekel}, A., {et~al.} 2020{\natexlab{b}},
  \mnras, 499, 2912, \dodoi{10.1093/mnras/staa2790}

\bibitem[{Gandolfi {et~al.}(2021)Gandolfi, Lapi, \&
  Liberati}]{Gandolfi:2021jai}
Gandolfi, G., Lapi, A., \& Liberati, S. 2021, Astrophys. J., 910, 76,
  \dodoi{10.3847/1538-4357/abe460}

\bibitem[{{Gandolfi} {et~al.}(2022){Gandolfi}, {Lapi}, \&
  {Liberati}}]{2022ApJ...929...48G}
{Gandolfi}, G., {Lapi}, A., \& {Liberati}, S. 2022, \apj, 929, 48,
  \dodoi{10.3847/1538-4357/ac5970}

\bibitem[{{Ghirardini} {et~al.}(2018){Ghirardini}, {Ettori}, {Eckert},
  {Molendi}, {Gastaldello}, {Pointecouteau}, {Hurier}, \&
  {Bourdin}}]{2018A&A...614A...7G}
{Ghirardini}, V., {Ettori}, S., {Eckert}, D., {et~al.} 2018, \aap, 614, A7,
  \dodoi{10.1051/0004-6361/201731748}

\bibitem[{Ghirardini {et~al.}(2018)Ghirardini, Ettori, Eckert, Molendi,
  Gastaldello, Pointecouteau, Hurier, \& Bourdin}]{Ghirardini:2017apw}
Ghirardini, V., Ettori, S., Eckert, D., {et~al.} 2018, Astron. Astrophys., 614,
  A7, \dodoi{10.1051/0004-6361/201731748}

\bibitem[{{Ghirardini} {et~al.}(2019){Ghirardini}, {Eckert}, {Ettori},
  {Pointecouteau}, {Molendi}, {Gaspari}, {Rossetti}, {De Grandi}, {Roncarelli},
  {Bourdin}, {Mazzotta}, {Rasia}, \& {Vazza}}]{2019A&A...621A..41G}
{Ghirardini}, V., {Eckert}, D., {Ettori}, S., {et~al.} 2019, \aap, 621, A41,
  \dodoi{10.1051/0004-6361/201833325}

\bibitem[{Green \& Moffat(2019)}]{Green:2019cqm}
Green, M.~A., \& Moffat, J.~W. 2019, Phys. Dark Univ., 25, 100323,
  \dodoi{10.1016/j.dark.2019.100323}

\bibitem[{Haridasu {et~al.}(2021)Haridasu, Karmakar, De~Petris, Cardone, \&
  Maoli}]{Haridasu:2021hzq}
Haridasu, B.~S., Karmakar, P., De~Petris, M., Cardone, V.~F., \& Maoli, R.
  2021.
\newblock \doarXiv{2111.01101}

\bibitem[{Heavens {et~al.}(2017{\natexlab{a}})Heavens, Fantaye, Mootoovaloo,
  Eggers, Hosenie, Kroon, \& Sellentin}]{Heavens:2017afc}
Heavens, A., Fantaye, Y., Mootoovaloo, A., {et~al.} 2017{\natexlab{a}}.
\newblock \doarXiv{1704.03472}

\bibitem[{Heavens {et~al.}(2017{\natexlab{b}})Heavens, Fantaye, Sellentin,
  Eggers, Hosenie, Kroon, \& Mootoovaloo}]{Heavens:2017hkr}
Heavens, A., Fantaye, Y., Sellentin, E., {et~al.} 2017{\natexlab{b}}, Phys.
  Rev. Lett., 119, 101301, \dodoi{10.1103/PhysRevLett.119.101301}

\bibitem[{Hogg \& Foreman-Mackey(2018)}]{Hogg:2017akh}
Hogg, D.~W., \& Foreman-Mackey, D. 2018, Astrophys. J. Suppl., 236, 11,
  \dodoi{10.3847/1538-4365/aab76e}

\bibitem[{Ivanov \& Liberati(2020)}]{Ivanov:2019iec}
Ivanov, D., \& Liberati, S. 2020, JCAP, 07, 065,
  \dodoi{10.1088/1475-7516/2020/07/065}

\bibitem[{Jeffreys(1961)}]{Jeffreys:1939xee}
Jeffreys, H. 1961, {The Theory of Probability}, Oxford Classic Texts in the
  Physical Sciences

\bibitem[{{Keller} \& {Wadsley}(2017)}]{2017ApJ...835L..17K}
{Keller}, B.~W., \& {Wadsley}, J.~W. 2017, \apjl, 835, L17,
  \dodoi{10.3847/2041-8213/835/1/L17}

\bibitem[{Lelli {et~al.}(2017)Lelli, McGaugh, Schombert, \&
  Pawlowski}]{Lelli:2017vgz}
Lelli, F., McGaugh, S.~S., Schombert, J.~M., \& Pawlowski, M.~S. 2017,
  Astrophys. J., 836, 152, \dodoi{10.3847/1538-4357/836/2/152}

\bibitem[{Lewis(2019)}]{Lewis:2019xzd}
Lewis, A. 2019.
\newblock \doarXiv{1910.13970}

\bibitem[{Li {et~al.}(2018)Li, Lelli, McGaugh, \& Schombert}]{Li:2018tdo}
Li, P., Lelli, F., McGaugh, S., \& Schombert, J. 2018, Astron. Astrophys., 615,
  A3, \dodoi{10.1051/0004-6361/201732547}

\bibitem[{{Li} {et~al.}(2018){Li}, {Lelli}, {McGaugh}, \&
  {Schombert}}]{2018A&A...615A...3L}
{Li}, P., {Lelli}, F., {McGaugh}, S., \& {Schombert}, J. 2018, \aap, 615, A3,
  \dodoi{10.1051/0004-6361/201732547}

\bibitem[{{{\L}okas} \& {Mamon}(2001)}]{2001MNRAS.321..155L}
{{\L}okas}, E.~L., \& {Mamon}, G.~A. 2001, \mnras, 321, 155,
  \dodoi{10.1046/j.1365-8711.2001.04007.x}

\bibitem[{{Ludlow} {et~al.}(2017){Ludlow}, {Ben{\'\i}tez-Llambay}, {Schaller},
  {Theuns}, {Frenk}, {Bower}, {Schaye}, {Crain}, {Navarro}, {Fattahi}, \&
  {Oman}}]{2017PhRvL.118p1103L}
{Ludlow}, A.~D., {Ben{\'\i}tez-Llambay}, A., {Schaller}, M., {et~al.} 2017,
  \prl, 118, 161103, \dodoi{10.1103/PhysRevLett.118.161103}

\bibitem[{{Meneghetti} {et~al.}(2020){Meneghetti}, {Davoli}, {Bergamini},
  {Rosati}, {Natarajan}, {Giocoli}, {Caminha}, {Metcalf}, {Rasia}, {Borgani},
  {Calura}, {Grillo}, {Mercurio}, \& {Vanzella}}]{2020Sci...369.1347M}
{Meneghetti}, M., {Davoli}, G., {Bergamini}, P., {et~al.} 2020, Science, 369,
  1347, \dodoi{10.1126/science.aax5164}

\bibitem[{{Milgrom}(1983)}]{1983ApJ...270..365M}
{Milgrom}, M. 1983, \apj, 270, 365, \dodoi{10.1086/161130}

\bibitem[{{Navarro}(2006)}]{2006aglu.confE..30N}
{Navarro}, J. 2006, in KITP Conference: Applications of Gravitational Lensing:
  Unique Insights into Galaxy Formation and Evolution, ed. L.~V.~E. {Koopmans},
  C.-P. {Ma}, B.~{Moore}, P.~{Schneider}, \& T.~{Treu}, 30

\bibitem[{Navarro {et~al.}(2017)Navarro, Ben\'\i{}tez-Llambay, Fattahi, Frenk,
  Ludlow, Oman, Schaller, \& Theuns}]{Navarro:2016bfs}
Navarro, J.~F., Ben\'\i{}tez-Llambay, A., Fattahi, A., {et~al.} 2017, Mon. Not.
  Roy. Astron. Soc., 471, 1841, \dodoi{10.1093/mnras/stx1705}

\bibitem[{Navarro {et~al.}(1996)Navarro, Frenk, \& White}]{Navarro:1995iw}
Navarro, J.~F., Frenk, C.~S., \& White, S. D.~M. 1996, Astrophys. J., 462, 563,
  \dodoi{10.1086/177173}

\bibitem[{{Peirani} {et~al.}(2017){Peirani}, {Dubois}, {Volonteri},
  {Devriendt}, {Bundy}, {Silk}, {Pichon}, {Kaviraj}, {Gavazzi}, \&
  {Habouzit}}]{2017MNRAS.472.2153P}
{Peirani}, S., {Dubois}, Y., {Volonteri}, M., {et~al.} 2017, \mnras, 472, 2153,
  \dodoi{10.1093/mnras/stx2099}

\bibitem[{{Planck Collaboration} {et~al.}(2016){Planck Collaboration}, {Ade},
  {Aghanim}, {Arnaud}, {Ashdown}, {Aumont}, {Baccigalupi}, {Banday},
  {Barreiro}, {Bartlett}, {Bartolo}, {Battaner}, {Battye}, {Benabed},
  {Beno{\^\i}t}, {Benoit-L{\'e}vy}, {Bernard}, {Bersanelli}, {Bielewicz},
  {Bock}, {Bonaldi}, {Bonavera}, {Bond}, {Borrill}, {Bouchet}, {Bucher},
  {Burigana}, {Butler}, {Calabrese}, {Cardoso}, {Catalano}, {Challinor},
  {Chamballu}, {Chary}, {Chiang}, {Christensen}, {Church}, {Clements},
  {Colombi}, {Colombo}, {Combet}, {Comis}, {Couchot}, {Coulais}, {Crill},
  {Curto}, {Cuttaia}, {Danese}, {Davies}, {Davis}, {de Bernardis}, {de Rosa},
  {de Zotti}, {Delabrouille}, {D{\'e}sert}, {Diego}, {Dolag}, {Dole},
  {Donzelli}, {Dor{\'e}}, {Douspis}, {Ducout}, {Dupac}, {Efstathiou}, {Elsner},
  {En{\ss}lin}, {Eriksen}, {Falgarone}, {Fergusson}, {Finelli}, {Forni},
  {Frailis}, {Fraisse}, {Franceschi}, {Frejsel}, {Galeotta}, {Galli}, {Ganga},
  {Giard}, {Giraud-H{\'e}raud}, {Gjerl{\o}w}, {Gonz{\'a}lez-Nuevo},
  {G{\'o}rski}, {Gratton}, {Gregorio}, {Gruppuso}, {Gudmundsson}, {Hansen},
  {Hanson}, {Harrison}, {Henrot-Versill{\'e}}, {Hern{\'a}ndez-Monteagudo},
  {Herranz}, {Hildebrandt}, {Hivon}, {Hobson}, {Holmes}, {Hornstrup}, {Hovest},
  {Huffenberger}, {Hurier}, {Jaffe}, {Jaffe}, {Jones}, {Juvela},
  {Keih{\"a}nen}, {Keskitalo}, {Kisner}, {Kneissl}, {Knoche}, {Kunz},
  {Kurki-Suonio}, {Lagache}, {L{\"a}hteenm{\"a}ki}, {Lamarre}, {Lasenby},
  {Lattanzi}, {Lawrence}, {Leonardi}, {Lesgourgues}, {Levrier}, {Liguori},
  {Lilje}, {Linden-V{\o}rnle}, {L{\'o}pez-Caniego}, {Lubin},
  {Mac{\'\i}as-P{\'e}rez}, {Maggio}, {Maino}, {Mandolesi}, {Mangilli}, {Maris},
  {Martin}, {Mart{\'\i}nez-Gonz{\'a}lez}, {Masi}, {Matarrese}, {McGehee},
  {Meinhold}, {Melchiorri}, {Melin}, {Mendes}, {Mennella}, {Migliaccio},
  {Mitra}, {Miville-Desch{\^e}nes}, {Moneti}, {Montier}, {Morgante},
  {Mortlock}, {Moss}, {Munshi}, {Murphy}, {Naselsky}, {Nati}, {Natoli},
  {Netterfield}, {N{\o}rgaard-Nielsen}, {Noviello}, {Novikov}, {Novikov},
  {Oxborrow}, {Paci}, {Pagano}, {Pajot}, {Paoletti}, {Partridge}, {Pasian},
  {Patanchon}, {Pearson}, {Perdereau}, {Perotto}, {Perrotta}, {Pettorino},
  {Piacentini}, {Piat}, {Pierpaoli}, {Pietrobon}, {Plaszczynski},
  {Pointecouteau}, {Polenta}, {Popa}, {Pratt}, {Pr{\'e}zeau}, {Prunet},
  {Puget}, {Rachen}, {Rebolo}, {Reinecke}, {Remazeilles}, {Renault}, {Renzi},
  {Ristorcelli}, {Rocha}, {Roman}, {Rosset}, {Rossetti}, {Roudier},
  {Rubi{\~n}o-Mart{\'\i}n}, {Rusholme}, {Sandri}, {Santos}, {Savelainen},
  {Savini}, {Scott}, {Seiffert}, {Shellard}, {Spencer}, {Stolyarov}, {Stompor},
  {Sudiwala}, {Sunyaev}, {Sutton}, {Suur-Uski}, {Sygnet}, {Tauber}, {Terenzi},
  {Toffolatti}, {Tomasi}, {Tristram}, {Tucci}, {Tuovinen}, {T{\"u}rler},
  {Umana}, {Valenziano}, {Valiviita}, {Van Tent}, {Vielva}, {Villa}, {Wade},
  {Wandelt}, {Wehus}, {Weller}, {White}, {Yvon}, {Zacchei}, \&
  {Zonca}}]{2016A&A...594A..24P}
{Planck Collaboration}, {Ade}, P.~A.~R., {Aghanim}, N., {et~al.} 2016, \aap,
  594, A24, \dodoi{10.1051/0004-6361/201525833}

\bibitem[{Pontzen \& Governato(2014)}]{Pontzen:2014lma}
Pontzen, A., \& Governato, F. 2014, Nature, 506, 171,
  \dodoi{10.1038/nature12953}

\bibitem[{Robertson {et~al.}(2017)Robertson, Massey, \&
  Eke}]{Robertson:2016xjh}
Robertson, A., Massey, R., \& Eke, V. 2017, Mon. Not. Roy. Astron. Soc., 465,
  569, \dodoi{10.1093/mnras/stw2670}

\bibitem[{Rodrigues \& Marra(2020)}]{rodrigues_marra_2020}
Rodrigues, D.~C., \& Marra, V. 2020, Proceedings of the International
  Astronomical Union, 15, 457–459, \dodoi{10.1017/S1743921320001684}

\bibitem[{{Rubin} {et~al.}(1978){Rubin}, {Ford}, \&
  {Thonnard}}]{1978ApJ...225L.107R}
{Rubin}, V.~C., {Ford}, W.~K., J., \& {Thonnard}, N. 1978, \apjl, 225, L107,
  \dodoi{10.1086/182804}

\bibitem[{{Salucci}(2019)}]{2019A&ARv..27....2S}
{Salucci}, P. 2019, \aapr, 27, 2, \dodoi{10.1007/s00159-018-0113-1}

\bibitem[{Salucci \& Burkert(2000)}]{Salucci:2000ps}
Salucci, P., \& Burkert, A. 2000, Astrophys. J. Lett., 537, L9,
  \dodoi{10.1086/312747}

\bibitem[{{Santos-Santos} {et~al.}(2016){Santos-Santos}, {Brook}, {Stinson},
  {Di Cintio}, {Wadsley}, {Dom{\'\i}nguez-Tenreiro}, {Gottl{\"o}ber}, \&
  {Yepes}}]{2016MNRAS.455..476S}
{Santos-Santos}, I.~M., {Brook}, C.~B., {Stinson}, G., {et~al.} 2016, \mnras,
  455, 476, \dodoi{10.1093/mnras/stv2335}

\bibitem[{Tian {et~al.}(2020)Tian, Umetsu, Ko, Donahue, \& Chiu}]{Tian:2020qjd}
Tian, Y., Umetsu, K., Ko, C.-M., Donahue, M., \& Chiu, I.-N. 2020, Astrophys.
  J., 896, 70, \dodoi{10.3847/1538-4357/ab8e3d}

\bibitem[{Trotta(2008)}]{Trotta:2008qt}
Trotta, R. 2008, Contemp. Phys., 49, 71, \dodoi{10.1080/00107510802066753}

\bibitem[{Trotta(2017)}]{Trotta:2017wnx}
Trotta, R. 2017, in {Bayesian Methods in Cosmology}.
\newblock \doarXiv{1701.01467}

\bibitem[{{Umetsu} {et~al.}(2014){Umetsu}, {Medezinski}, {Nonino}, {Merten},
  {Postman}, {Meneghetti}, {Donahue}, {Czakon}, {Molino}, {Seitz}, {Gruen},
  {Lemze}, {Balestra}, {Ben{\'\i}tez}, {Biviano}, {Broadhurst}, {Ford},
  {Grillo}, {Koekemoer}, {Melchior}, {Mercurio}, {Moustakas}, {Rosati}, \&
  {Zitrin}}]{2014ApJ...795..163U}
{Umetsu}, K., {Medezinski}, E., {Nonino}, M., {et~al.} 2014, \apj, 795, 163,
  \dodoi{10.1088/0004-637X/795/2/163}

\bibitem[{Vikhlinin {et~al.}(2006)Vikhlinin, Kravtsov, Forman, Jones,
  Markevitch, Murray, \& Van~Speybroeck}]{Vikhlinin:2005mp}
Vikhlinin, A., Kravtsov, A., Forman, W., {et~al.} 2006, Astrophys. J., 640,
  691, \dodoi{10.1086/500288}

\bibitem[{Wheeler {et~al.}(2019)Wheeler, Hopkins, \& Doré}]{Wheeler_2019}
Wheeler, C., Hopkins, P.~F., \& Doré, O. 2019, The Astrophysical Journal, 882,
  46, \dodoi{10.3847/1538-4357/ab311b}

\bibitem[{{Zwicky}(1933)}]{1933AcHPh...6..110Z}
{Zwicky}, F. 1933, Helvetica Physica Acta, 6, 110

\end{thebibliography}
\bibliographystyle{aasjournal}

\clearpage

\begin{deluxetable*}{ccccc}
\tablecaption{Reduced $\chi^2$ from the MCMC parameter estimation for both the NFW and NMC DM models alongside the Bayesian evidence $\Delta_{\mathcal{B}}$ in favor of the NMC DM model. \label{chisq} }
\tablewidth{0pt}
\tablehead{
\colhead{Cluster} & \colhead{z} & \colhead{$\chi^2_{\mathrm{red, NFW}}$} & \colhead{$\chi^2_{\mathrm{red, NMC}}$} & \colhead{$\Delta_{\mathcal{B}}$}
}
\startdata
\vspace{-0.35cm} &&&\\
\vspace{0.0cm} A85 & 0.0555 & 2.9 & 2.7 & -0.89 \\
\vspace{0.0cm} A644 & 0.0704 & 2.4 & 2.2 & 0.11 \\
\vspace{0.0cm} A1644 & 0.0473 & 3.9 & 3.4 & 1.01 \\
\vspace{0.0cm} A1759 & 0.0622 & 1.7 & 1.6 & 1.34 \\
\vspace{0.0cm} A2029 & 0.0773 & 1.6 & 1.6 & -0.15  \\
\vspace{0.0cm} A2142 & 0.0909 & 3.3 & 3.3 & -1.32 \\
\vspace{0.0cm} A2255 & 0.0809 & 6.7 & 1.8 & 2.64 \\
\vspace{0.0cm} A2319 & 0.0557 & 7.8 & 7.1 & 2.05 \\
\vspace{0.0cm} A3158 & 0.0597 & 2.3 & 2.1 & 2.81 \\
\vspace{0.0cm} A3266 & 0.0589 & 6.7 & 6.8 & -1.89 \\
\vspace{0.0cm} RXC1825 & 0.0650 & 3.3 & 6.1 & -3.53 \\
\vspace{0.0cm} ZW1215 & 0.0766 & 0.97 & 0.86 & -0.81 \\
\vspace{-0.35cm} &&&\\
\enddata
\end{deluxetable*}

\begin{deluxetable*}{cccccc}
\tablenum{2}
\tablecaption{Results of the MCMC parameter estimation for the NFW models and the NMC DM model. The full set of parameter values and the related contours are available in the figure set. \label{fit}}
\tablewidth{0pt}
\tablehead{
\colhead{Cluster} & \colhead{$c_{500, \mathrm{GR}}$} & \colhead{$M_{500, \mathrm{GR}}/(10^{14} M_{\odot})$} & \colhead{$c_{500, \mathrm{NMC}}$} & \colhead{$M_{500, \mathrm{NMC}}/(10^{14} M_{\odot})$} & \colhead{$\Lnmc$ [kpc]}
}
\startdata
\vspace{-0.35cm} &&&\\
\vspace{0.0cm} A85 & $2.0^{+0.1}_{-0.1}$ & $6.2^{+0.2}_{-0.3}$ & $2.0 \pm 0.2$ & $6.2^{+0.3}_{-0.3}$ & $12^{+6}_{-10}$ \\
\vspace{0.0cm} A644 & $5^{+1}_{-2}$ & $4.6\pm 0.4$ & $ 6.2^{+1.0}_{-1.7}$ & $4.6 \pm 0.4$ & $26^{+11}_{-4}$ \\
\vspace{0.0cm} A1644 & $1.1^{+0.2}_{-0.4}$ & $3.2^{+0.3}_{-0.4}$ & $1.4^{+0.3}_{-0.5}$ & $3.3^{+0.3}_{-0.3}$ & $27^{+9}_{-3}$ \\
\vspace{0.0cm} A1759 & $3.0\pm 0.2$ & $4.7^{+0.2}_{-0.3}$ & $3.4^{+0.3}_{-0.3}$ & $4.6^{+0.2}_{-0.2}$ & $20^{+7}_{-3}$ \\
\vspace{0.0cm} A2029 & $3.3^{+0.2}_{-0.3}$ & $7.7 \pm 0.4$ & $3.6^{+0.3}_{-0.5}$ & $7.5\pm 0.4$ & $28^{+15}_{-7}$ \\
\vspace{0.0cm} A2142 & $2.3^{+0.2}_{-0.2}$ & $8.4\pm 0.4$ & $2.4^{+0.2}_{-0.3}$ & $8.4^{+0.4}_{-0.4}$ & $17.3^{+7}_{-15}$ \\
\vspace{0.0cm} A2255 & $1.6^{+0.4}_{-0.9}$ & $4.7\pm 0.4$ & $2.3^{+0.3}_{-0.7}$ & $4.8\pm 0.3$ & $109^{+11}_{-8}$ \\
\vspace{0.0cm} A2319 & $3.8^{+0.4}_{-0.6}$ & $7.4\pm 0.2$ & $4.6^{+0.6}_{-0.8}$ & $7.4 \pm 0.2$ & $60^{+17}_{-5}$ \\
\vspace{0.0cm} A3158 & $2.0^{+0.3}_{-0.4}$ & $4.0^{+0.3}_{-0.3} $ & $2.6^{+0.4}_{-0.4}$ & $4.0 \pm 0.2$ & $36^{+6}_{-3}$ \\
\vspace{0.0cm} A3266 & $1.7^{+0.2}_{-0.2}$ & $6.6\pm 0.2$ & $1.7\pm 0.2$ & $6.5^{+0.3}_{-0.3}$ & $15^{+5}_{-13}$ \\
\vspace{0.0cm} RXC1825 & $2.6^{+0.4}_{-0.4}$ & $4.1^{+0.3}_{-0.3}$ & $4.0^{+0.5}_{-0.9}$ & $3.5 \pm 0.3$ & $9^{+3}_{-8}$ \\
\vspace{0.0cm} ZW1215 & $1.5^{+0.2}_{-0.3}$ & $7.1\pm 0.7$ & $1.6^{+0.2}_{-0.3}$ & $7.0 \pm 0.6$ & $22^{+9}_{-19}$ \\
\vspace{-0.35cm} &&&\\
\enddata
\end{deluxetable*}

\clearpage

\appendix
\section{Pressure profiles}
In this section, we show the pressure profiles reconstructed for the NMC DM case for each of the 12 clusters. 

\begin{figure}
    
    \includegraphics[scale= 0.2]{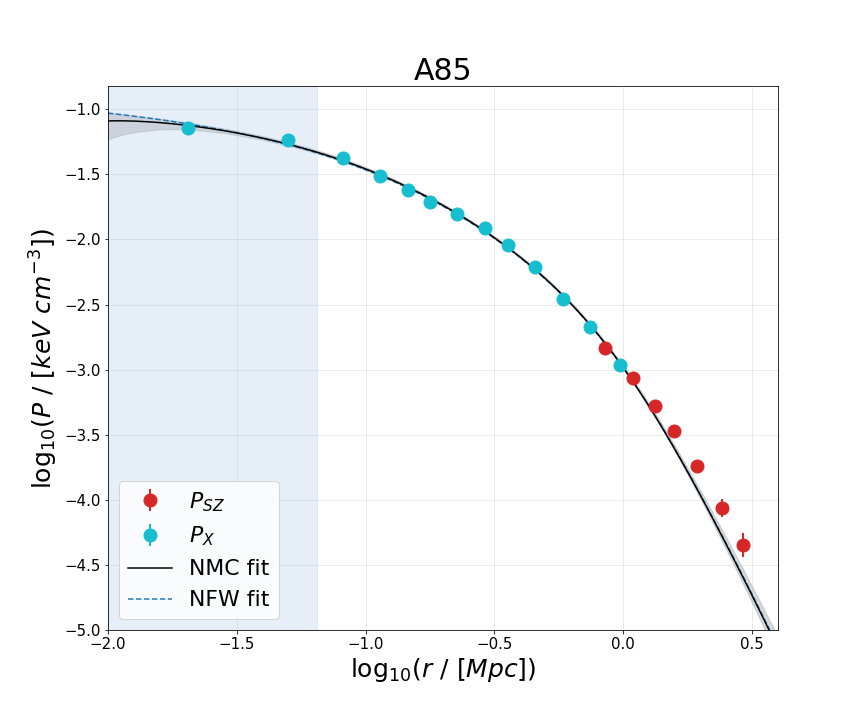}
    \includegraphics[scale= 0.2]{A644.png}
    \includegraphics[scale= 0.2]{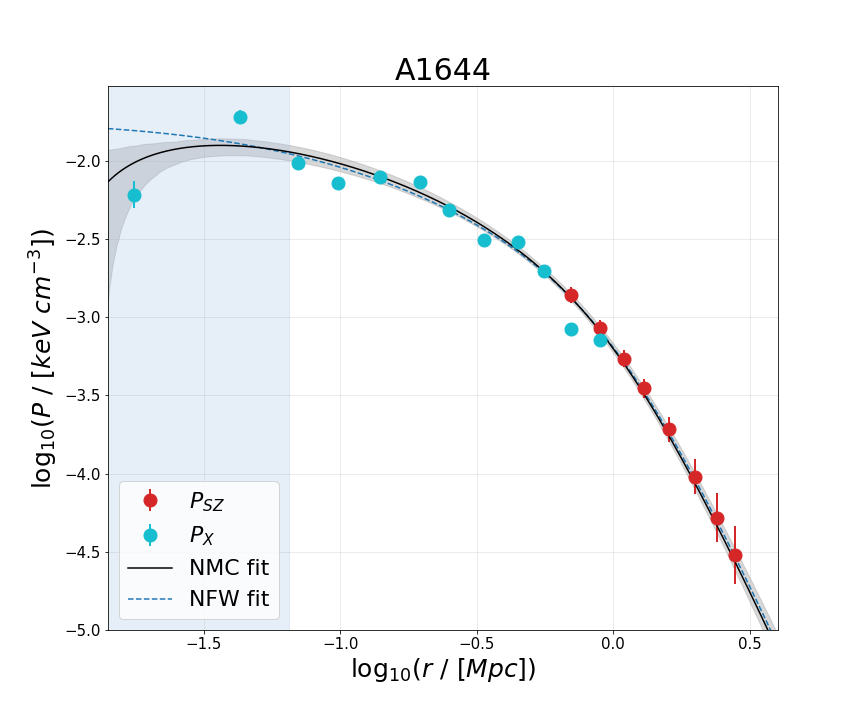}
    \vfill
    \includegraphics[scale= 0.2]{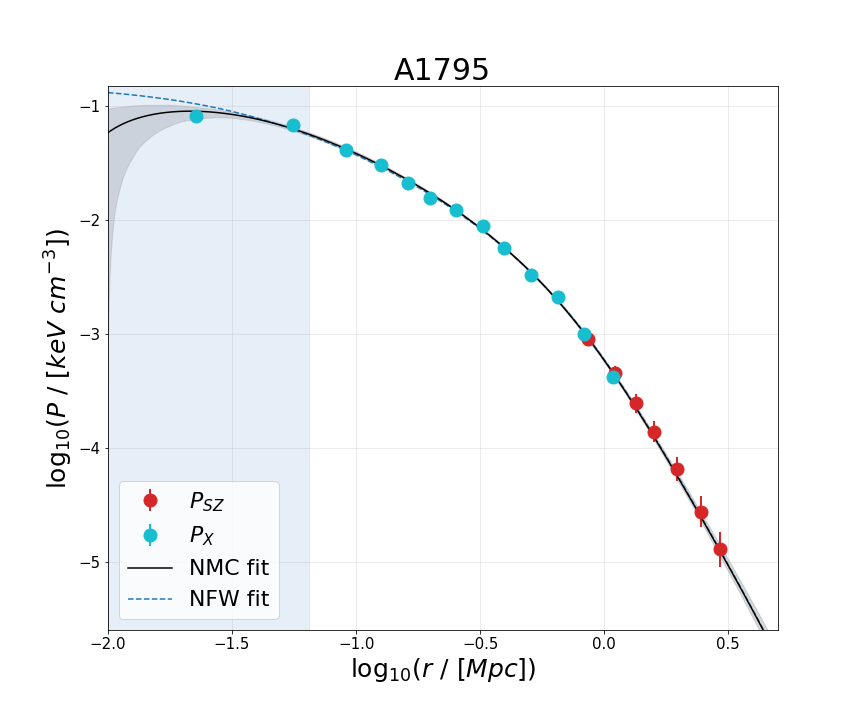}
    \includegraphics[scale= 0.2]{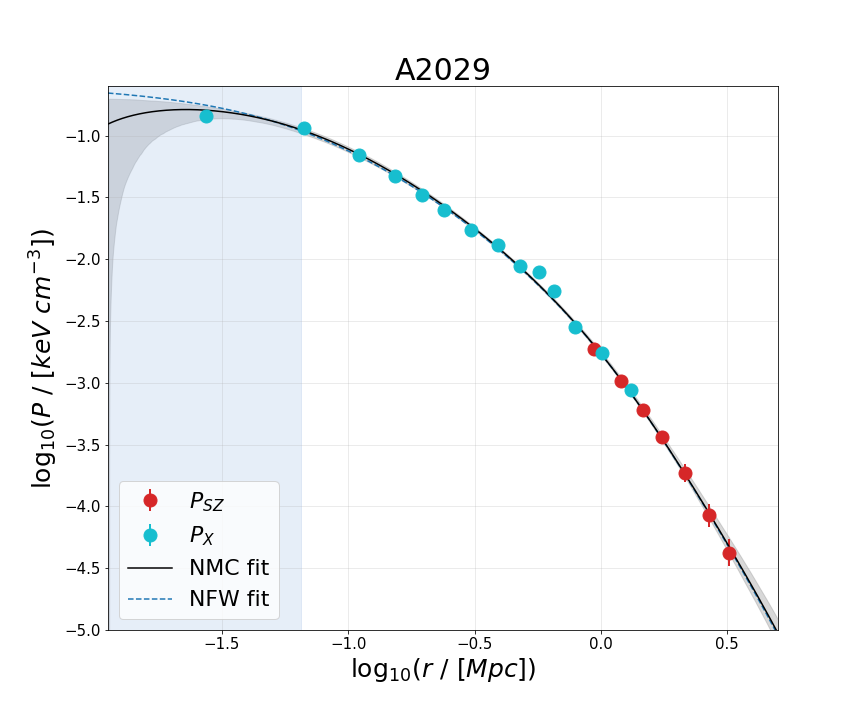}
    \includegraphics[scale= 0.2]{A2142.png}
    \vfill
    \includegraphics[scale= 0.2]{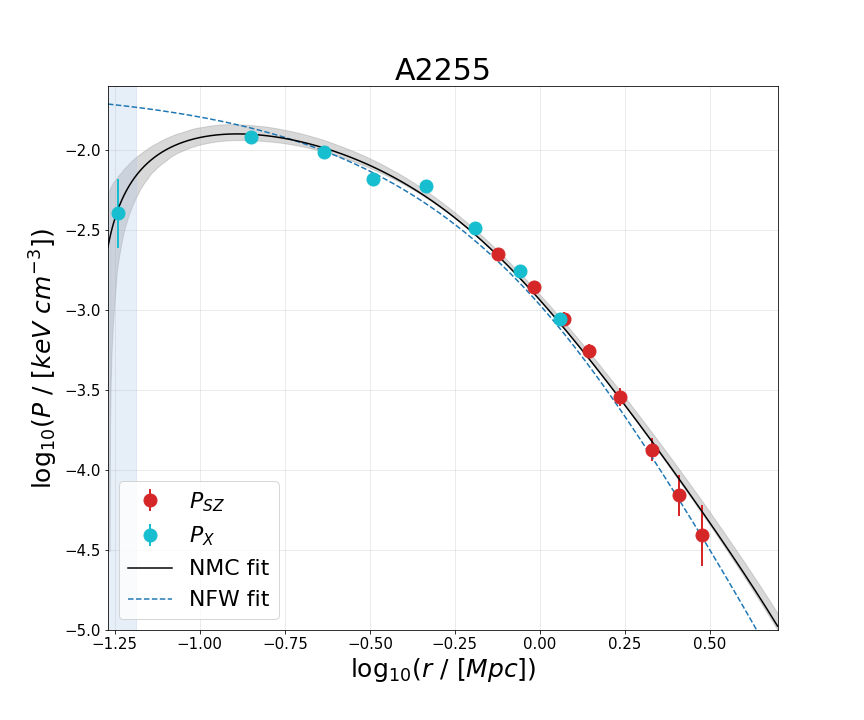}
    \includegraphics[scale= 0.2]{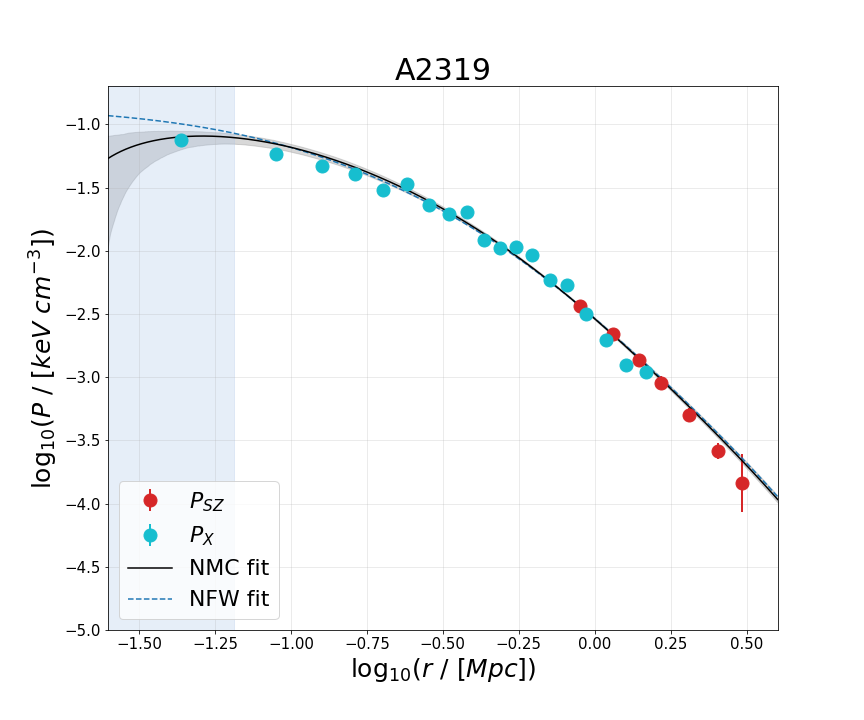}
    \includegraphics[scale= 0.2]{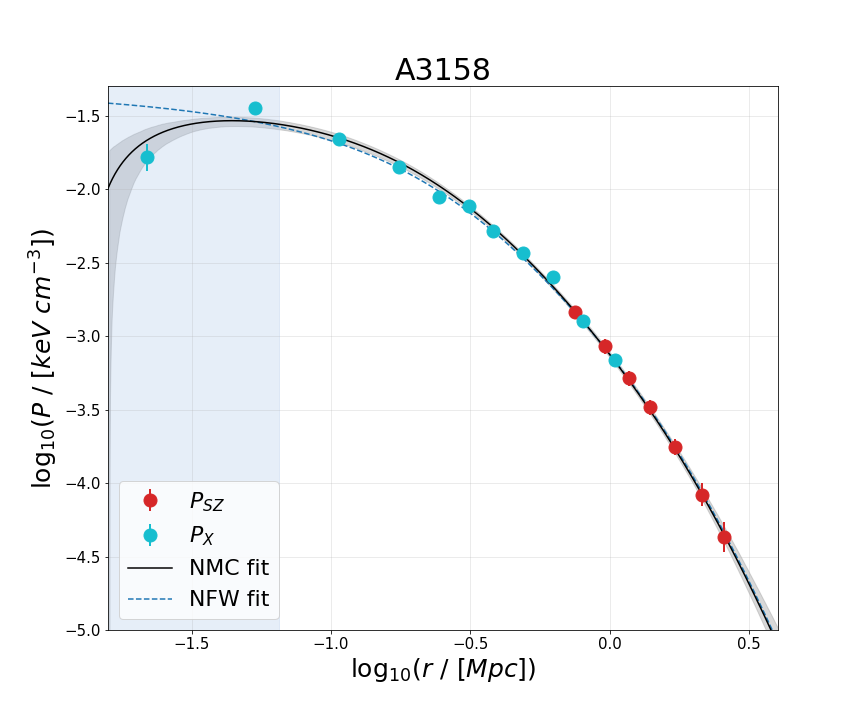}
    \vfill
    \includegraphics[scale= 0.2]{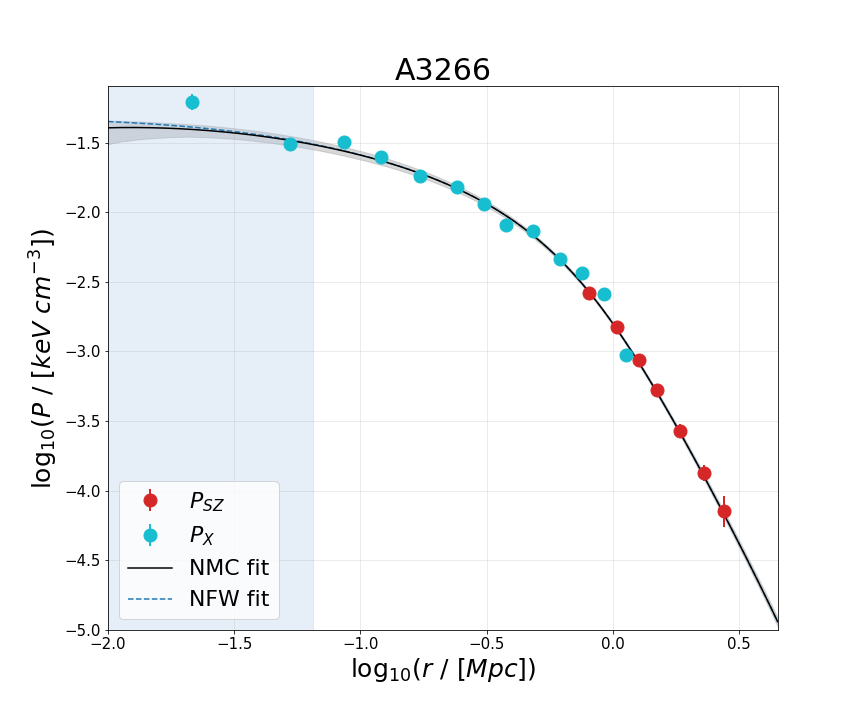}
    \includegraphics[scale= 0.2]{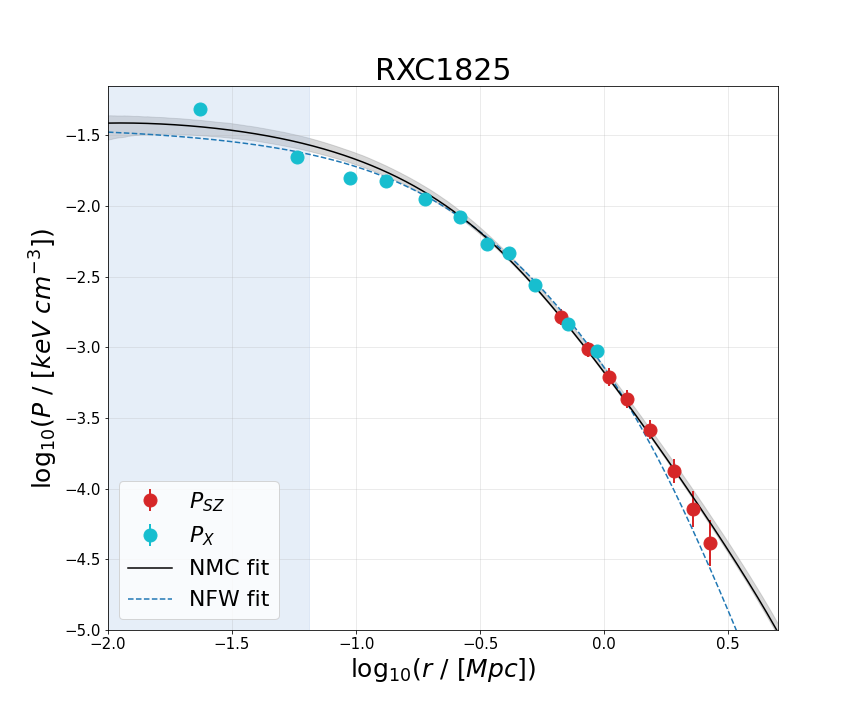}
    \includegraphics[scale= 0.2]{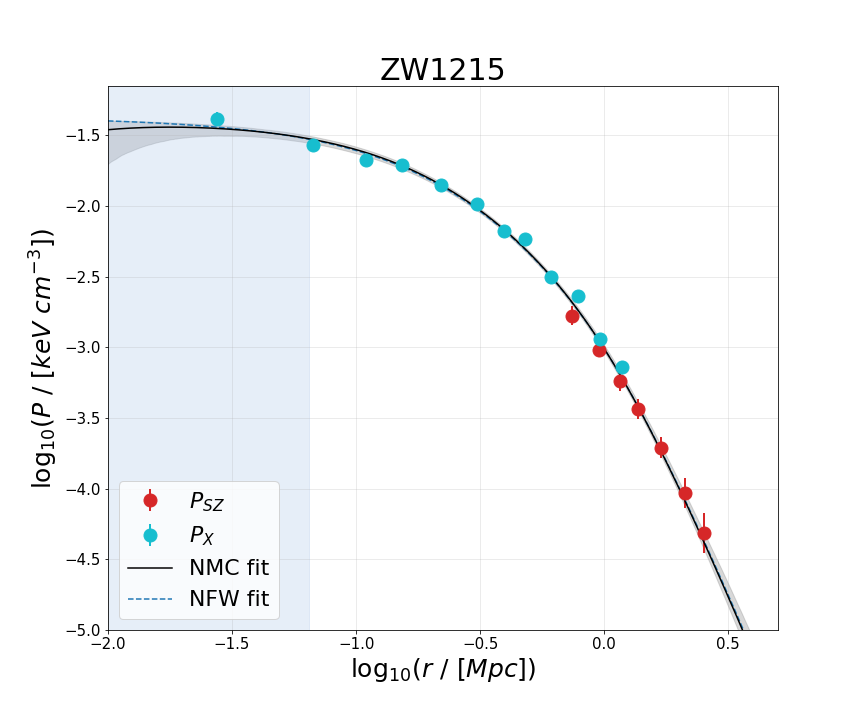}
    \caption{We show the comparison of the pressure profiles for the NMC DM model (solid lines), against those of the NFW model (dashed grey lines) for all the 12 clusters in X-COP compilation.  }
    \label{fig:profiles_all}
\end{figure}

%



\end{document}